\documentclass[fleqn,usenatbib,useAMS]{mnras}

\usepackage{newtxtext,newtxmath}

\usepackage[T1]{fontenc}

\DeclareRobustCommand{\VAN}[3]{#2}
\let\VANthebibliography\thebibliography
\def\thebibliography{\DeclareRobustCommand{\VAN}[3]{##3}\VANthebibliography}

\usepackage{graphicx}	
\usepackage{amsmath}	
\usepackage{orcidlink}

\title[Radio emission from flickering AGN jets]{Simulating radio emission from flickering AGN jets: travelling shocks and hotspot brightening}

\author[E. L. Elley et al.]{
Emma L. Elley$^{\orcidlink{0009-0002-5349-908X},1}$\thanks{E-mail: emma.elley@physics.ox.ac.uk}, 
James H. Matthews$^{\orcidlink{0000-0002-3493-7737},1}$, 
Dipanjan Mukherjee$^{\orcidlink{0000-0003-0632-1000},2}$, 
and Bhargav Vaidya$^{\orcidlink{0000-0001-5424-0059},3}$
\\
$^{1}$Astrophysics Subdepartment, Department of Physics, University of Oxford, Keble Road, Oxford, OX13RH, UK\\
$^{2}$Inter-University Centre for Astronomy \& Astrophysics, Post Bag 4, Ganeshkhind, Pune 411007, India\\
$^{3}$Department of Astronomy, Astrophysics and Space Engineering, Indian Institute of Technology, Indore 452020, India.
}

\date{Accepted 2026 January 16. Received 2026 January 06; in original form 2025 September 23}

\pubyear{\the\year{}}

\begin{document}
\label{firstpage}
\pagerange{\pageref{firstpage}--\pageref{lastpage}}
\maketitle

\begin{abstract}
We investigate the impact of flickering variability in jet power on the luminosity and morphology of radio galaxies. We use a Lagrangian particle method together with relativistic hydrodynamics simulations using the PLUTO code to track the evolution of electron spectra through particle acceleration at shocks and cooling processes. We introduce an adapted version of this method which improves tracking of adiabatic cooling in regimes where low density jet material mixes with high density from the ambient medium in the lobes. We find that rapid increases in jet power can lead to large increases in hotspot luminosity due to the interaction of a travelling shock structure with the pre-existing shock structure at the jet head. We show that in some cases it may be possible to identify a bright region of emission corresponding to a shock travelling along the jet axis. We find that the time-averaged radiative efficiency of variable jets is similar to their steady counterparts, but find significant departures from this on an instantaneous basis. We suggest that, together with environmental effects and differences in the average powers of jets, variable jet powers may have a significant impact on how we understand the diversity of radio jets seen in observations and have significant implications for interpretations of jet powers, energy budgets and luminosity-linear size diagrams.
\end{abstract}

\begin{keywords}
 galaxies: jets -- radio continuum: galaxies -- acceleration of particles -- radiation mechanisms: non-thermal -- hydrodynamics -- shock waves
\end{keywords}



\section{Introduction}
\label{sec:intro}

Radio galaxies are observed over a wide range of scales and frequencies. In the radio, it is often possible to resolve the kpc scale structures of the radio lobes, showing structures such as emission along the jet line, hot spots, backflows and extended emission in the cocoon area. One example, Cygnus A, has been well-studied at multiple frequencies over a long period of time, including observations using the Very Large Array (VLA)~\citep{Carilli1991MultifrequencyGalaxies} and more recently with the Low Frequency Array (LOFAR)~\citep{McKean2016LOFARSpectra}. This source is an example of an FRII radio galaxy with the brightest regions (the hotspots) occurring further from the galaxy~\citep{Carilli1991MultifrequencyGalaxies}. These observations can provide important information about the behaviour of these jets. However, as they evolve over millions of years, each individual source reveals only a snapshot into the evolutionary process. Simulations are a valuable tool with the potential to bridge the gap between ideas about behaviour over longer timescales to the instantaneous view we get from observations. Here we seek to investigate how the consequences of variable fuelling of the accretion disk might affect radio observations over kpc scales. We link this to ideas about FRI and FRII morphologies~\citep{Fanaroff1974TheLuminosity}.

To sustain an AGN accretion disk a mechanism is needed to bring material toward the centre, linking the central observable hot accretion disk to what is happening in the wider galaxy. A supermassive black hole (SMBH) is not thought to be active for its full lifetime~\citep[e.g.][and references therein]{Morganti2017ArchaeologySpectrum}. In particular radio and optical observational studies have led to the conclusion that SMBHs can show recurrent behaviour~\citep{Saikia2009RecurrentNuclei}, with efforts being dedicated to estimating jet lifetimes~\citep[e.g.][]{Shabala2008TheGalaxies,Vantyghem2014CyclingClusters,Harwood2015SpectralGalaxies} in order to constrain the duty cycle of AGN. Radio lobes with double-double morphologies~\citep[e.g.][]{2000SchoenmakersRadioNuclei,2007BrocksoppThreeGalaxy,2019MahatmaLoTSSField} are often explained by restarting jets and these have been studied in simulations~\citep[e.g.][]{1991ClarkeNumericalJet, 2020WalgRelativisticGalaxies}. The variability and lifetimes of active periods are thought to have different characteristics based on the environment and mode of AGN activity. In the case of fuelling by dense, cold gas a radiatively efficient disk forms, where the cold gas in the nuclear region is thought to have been transported towards the central region of the galaxy primarily through secular processes~\citep{Heckman2014TheUniverse}, but may also be the result of galaxy mergers~\citep[e.g.][]{RamosAlmeida2012AreInteractions}. 
In more massive bulges and elliptical galaxies, which host more massive black holes, fuelling is thought to involve hot gas accretion~\citep{Heckman2014TheUniverse}. Galaxies undergoing powerful jet-mode feedback are found in dense galaxy group and clusters, which often have strong cooling flows~\citep{Heckman2014TheUniverse} as a result of short radiative cooling times, with the amount of cooling regulated by AGN feedback~\citep{Fabian2023HiddenSample}.

Building on earlier work by \cite{king2006}, \cite{Gaspari2013ChaoticHoles} introduced the theory of chaotic cold accretion (CCA), in which chaotic accretion episodes are driven by large scale hot gas undergoing cooling, heating and stirring. CCA provides a mechanism by which hot gas does not need to be directly accreted via Bondi accretion, but instead cools prior to being accreted. This work used hydrodynamic simulations with adaptive mesh refinement to study the fuelling of the central accretion disk from 50kpc to sub-pc scales, and found that clouds and filaments condense from the gas and rain onto the central region. CCA predicts a lognormal distribution in accretion rate with pink noise~\citep{Gaspari2017RainingModel}. Should the accretion rate vary in this way, it would be expected that the jet power would react to this variation. Whilst the disk-jet connection remains an active area of research, many general relativistic magnetohydrodynamic (GRMHD) simulations of accretion disks have naturally led to the formation of jets, including both accretion onto black holes~\citep[e.g.][]{Moscibrodzka2016General87,Dihingia2021Jetshole} and onto neutron stars~\citep[e.g.][]{Das2024Three-dimensionalJets}. The two leading theories describing the launching of jets from black holes are by~\cite{Blandford1982HydromagneticJets} and~\cite{Blandford1977ElectromagneticHoles}. The Blandford-Payne mechanism postulates jets launched by the removal of angular momentum from accretion disks via magnetic field lines. In contrast, the Blandford-Znajek mechanism describes jets launched by the frame-dragging of magnetic field lines by the spin of the black hole. The Blandford-Znajek mechanism predicts a jet power that is proportional to the mass accretion rate~\citep{Blandford1977ElectromagneticHoles} (see Section~\ref{sec:jet_powers}). Assuming CCA as a fuelling mechanism and that the BZ mechanism acts to launch the jet, we would also expect a noisy jet power with a lognormal distribution.

 Once a jet has been launched, it typically undergoes a series of recollimation shocks as it interacts with the cocoon region. The use of 3D simulations rather than 2D is crucial, particularly in situations where there is direct interest in shock structures and strength along the jet axis. Recent work has shown that recollimation and reflection shocks exhibit significantly different behaviours in 2D axisymmetric and 3D simulations, with 2D axisymmetric simulations and analysis predicting a series of strong recollimation and reflection shocks~\citep[e.g.][]{Komissarov1997SimulationsSources}, which are not reproduced in 3D simulations due to the effects of instabilities~\citep[e.g.][]{Massaglia2016MakingJets, Costa2025HowJets}.

After travelling the length of the jet, material reaches the boundary with the ambient medium. \cite{Blandford1974ASources} described the working surface at the head of a jet. At this surface material which has passed though a shock at the end of the beam meets material which has passed through the bow shock of the system in a contact discontinuity. The jet is confined in this region by the ram pressure of the intergalactic medium (IGM). The need for 3D simulations is further motivated by differences in the behaviour of turbulence and its effects of shock strengths at the head of the jet. ~\cite{Massaglia2016MakingJets} show that for jets with a density ratio of $\eta=10^{-2}$ those with a power below $\approx10^{43}$ erg s$^{-1}$ will dissipate energy via turbulence in 3D simulations, preventing the formation of a jet head and leading to an FRI-like morphology. This behaviour is not observed in 2D simulations, thus significantly altering the behaviour of the shock structure at the jet head.

Hydrodynamic simulations can be used to construct detailed models of the dynamics of jets and their interactions with surroundings. To link these with observations, it is useful to predict the emission expected from the systems. Because radio observations can provide a detailed, well spacially-resolved view of many radio jets, there has been much work to combine modelling of radio emission in particular with hydrodynamic simulations. A simple approach is to assume an emissivity based on pressure~\citep[e.g.][]{Longair2011HighAstrophysics,Hardcastle2013NumericalEnvironments} and a jet tracer. This method was used by~\cite{Whitehead2023StudyingJets} to study jets with powers varying according to a lognormal distribution of jet powers and a pink noise power spectral density. More complex methods aim to include information about particle acceleration at shocks more directly. An early example of this can be found in~\cite{Tregillis2004SyntheticAnalysis}, which introduced methods to predict radio and X-ray emission from simulations and analysed the application of observational techniques to these simulated emissivity maps. More recently the PRAiSE method~\citep{Yates-Jones2022PRAiSE:Sources} used Lagrangian tracer particles to capture information about the paths and pressures that populations of electrons follow. This data is then used to estimate an emissivity in post-processing using a semi-analytic model based on RAiSE II~\citep{Turner2018RAiSEAGN} where each population is assumed to have a power-law distribution of energies with an evolving maximum cut-off energy. \cite{Turner2018RAiSEAGN} showed that mixing of electrons of different ages changes the predicted spectrum of radio emission across a radio map for FRII sources, highlighting the importance of accounting for mixing in any method to predict radio emissivities from hydrodynamical simulations. A different approach to using Lagrangian particles is taken by the Lagrangian particle module~\citep{Vaidya2018AFlows} for the PLUTO code~\citep{Mignone2007PLUTO:Astrophysics}, where a binned spectrum of electron energies is evolved in tandem with relativistic hydrodynamics (RMHD) simulations. \cite{Dubey2024ParticlesSignatures} study steady jets and jets with a sinusoidally varying power using this module to predict the intensity of emission from the jet. They show that knots (compact regions of bright emission) proceed with an approximately constant velocity in both the case of a constant power jet and one with a sinusoidally variable jet speed.

Variability in observable features has also been seen in simulations of constant power jets.~\cite{Saxton2002ComplexA} use axisymmetric simulations to study the temporal variation of hotspots in steady jets, which they find to be described by a white noise spectrum on timescales longer than the dynamical timescale $t_{\rm dyn}\equiv 2r_j/c_{s,j}$, where $c_{s,j}$ is the speed of sound in the jet and $r_j$ is the jet radius. This dynamical timescale is shorter than the jet travel time from jet base to hotspot, leading the authors to the conclusion that the hotspots are causally disconnected and such that intrinsic variability in their brightnesses could affect estimations of quantities based on the ratio between the two brightnesses, e.g. relativistic beaming, with some simulations giving a probability of $> 20\%$ for a 4:1 or greater hotspot ratio. \cite{Saxton2010Time-dependentJets} highlight the importance of interactions between the jet and the ambient medium in determining the jet emission, arguing that this underscores the importance of accounting for environmental factors, whilst \citep{2022PeruchoLongEnvironments} highlight the dependence of the morphological features of constant power jets on properties of the ambient medium.

In this work, we use numerical simulations to explore the connection between jet power variation/flickering and the structure, morphology and dynamics of the resulting jet beam and hotspot. In the process, we explore (i) whether there exist morphological signatures of jet power variation in the large-scale radio observations of jets on scales up to 60 kpc, and (ii) how the luminosity and radiative efficiency of the system responds to variability. These aims are motivated by the predictions of CCA and wider goals of connecting radio morphologies to ideas about fuelling of accretion disks and radio jets over Myr timescales. Radio observations of kpc-scales provide an instantaneous view of radio jets, and these sources do not evolve on human timescales: the time for a disturbance to travel down a jet of length 30 kpc is at least $\approx0.1$ Myr (as bounded by the speed of light). However, this also means that even an instantaneous snapshot of a jet and its associated hotspot contains information about the input to the jet over the last fraction of a Myr. More generally, simulations provide a way to study the evolution of these systems on Myr timescales, providing an important tool to support the interpretation of both individual radio galaxy observations and the study of populations of radio galaxies. Specifically, we show herein that jet variability can produce travelling shock structures, dramatic short-term brightenings and changes to the brightness distributions of the jet-hotspot system. In addition to the primary limitation of our work -- long timescales involved -- there are various other caveats, which we discuss further in section~\ref{sec:limitations}; for example, we make the assumption throughout that a larger jet power will lead to a faster jet on kpc scales and that effects which occur in the launching and acceleration zone of the jet will not significantly impact this correlation. Throughout, our use of the Lagrangian particle module allows us to directly link energy injection into non-thermal electron populations through particle acceleration at shocks in the jet to the synchrotron radiation we see in radio observations. 

The paper is structured as follows. In Section~\ref{sec:methods} we describe the relativistic hydrodynamics simulation setup followed by the use of Lagrangian tracer particles. In Section~\ref{sec:results} we discuss the results from our constant and variable jet simulations, before discussing the results of jets with a step change in jet power. In Section~\ref{sec:discussion} we discuss the implications of our results for interpreting observations of radio galaxies, together with the limitations of our study and areas for future work. We conclude in Section~\ref{sec:conclusions}.

\section{Methods}
\label{sec:methods}

\begin{figure*}
    \centering
    \includegraphics[width=\linewidth]{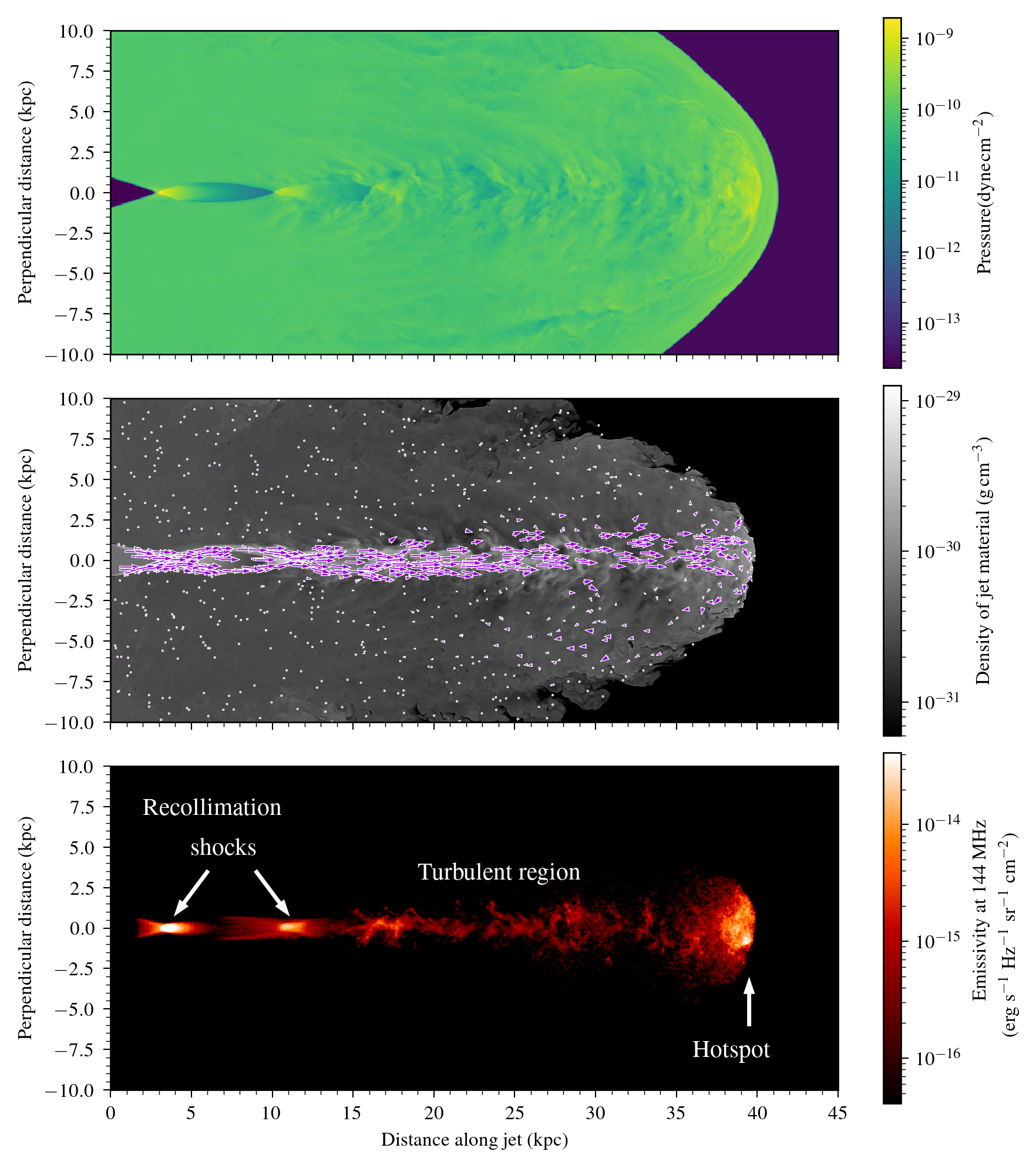}
    \caption{Emissivity map, pressure and density of jet material for a portion of the $\sigma=0.33$, seed 12 simulation at 9 Myr. The pressure and density maps show slices through the centre of the jet, whilst the emissivity map is summed along a line of sight perpendicular to the jet axis. The full simulation domain extends to 60 kpc in the jet direction and between -30 and 30 kpc in the two directions perpendicular to this. The locations and velocities of a small random sample of Lagrangian particles from the central slice through the domain are shown overlaid on the density of jet material. Velocities are shown as arrows where the length of the arrow scales linearly with speed.}
    \label{fig:method_summary}
\end{figure*}

In this work we model the radio emission from astrophysical jets with varying powers. We combine grid-based relativistic hydrodynamic (RHD) simulations using the PLUTO code~\citep{Mignone2007PLUTO:Astrophysics}, with test particle methods, introducing a novel approach to account for mixing between shocked and un-shocked material. We begin by describing the underlying setup. In Section~\ref{sec:jet_powers}, we discuss our approach to modelling variability in jet power in the simulations. In further sections, we discuss our use of the Lagrangian particle module for modelling non-thermal emission~\citep{Vaidya2018AFlows} of the PLUTO code. We describe how we inject the particles into the simulation in Appendix~\ref{sec:app_particle_injection}. This module allows the user to track the evolution of electron energy spectra in the jet through particle acceleration at shocks (Section~\ref{sec:shock_acceleration}), adiabatic cooling (Section~\ref{sec:tracer_method_derivation}) and synchrotron cooling (Section~\ref{sec:synchrotron_cooling}) and to calculate their emissivities.

\subsection{Simulation setup}
\label{sec:setup}

We run 3D RHD simulations on a $1500^3$ Cartesian grid extending 60 kpc in each direction, leading to a grid resolution of 0.04 kpc. We use RHD simulations as opposed to RMHD in order to separate the effects of variability in jet power from those of specific magnetic field geometries. We plan to investigate the effects of various magnetic field geometries using RMHD simulations in future work and discuss potential effects of this choice in Section~\ref{sec:limitations}. We show an example of our estimated fields in Appendix~\ref{sec:app_estimated_magnetic_field}.

The background density is set to a King profile where
\begin{equation}
    \rho(r) = \rho_0\left[1+\left(\frac{r}{r_c}\right)^2\right]^{-\frac{3\beta}{2}}
\end{equation}
with $r_{\rm c}=50$ kpc, $\beta = 0.5$ and $\rho_0=6.0\times10^{-27}$ g cm$^{-3}$ in keeping with the methods of~\cite{Whitehead2023StudyingJets}. We add small perturbations to this density in order to break the symmetry in the simulation.
We focus on the kpc scale behaviour and therefore do not model the jet launching region. We inject a jet with constant density $\rho_{\rm j}=10^{-4}\rho_0$ through a cylindrical nozzle. We apply outflow boundary conditions on all surfaces except the surface through which the jet enters, where we apply a reflective boundary condition outside of the jet radius ($r_{\rm j} = 1$ kpc). Within the jet radius we set the jet velocity, $v_{\rm j}$ to give the desired Lorentz factor, $\Gamma_{\rm j}$. We describe the calculation of this Lorentz factor in Section~\ref{sec:jet_powers}. We use the Harten, Lax, Van Leer Riemann solver with middle contact discontinuity restoration (HLLC) with a Courant-Friedrichs-Levy number of 0.3, linear reconstruction and 2nd-order Runge-Kutta time stepping. Simulation units are given in table~\ref{tab:sim_units}.

The top panel of Figure~\ref{fig:method_summary} shows the pressure in a central slice through the central region of the simulation. The jet inlet region is visible around (0,0) in the slice as a conical region. We discuss the structures seen in this figure in further detail in Section~\ref{sec:results}. We chose a resolution of 0.04 kpc in each direction, as we found this was required to ensure that the distribution of the Lagrangian particles sufficiently tracks the distribution of the fluid tracer in the backflow and cocoon regions (see Appendix~\ref{sec:particles_fluid_tracer}). This results in a physically large jet radius compared to what might be expected from observations, but is a necessary choice to reach 50 kpc scales whilst ensuring the Lagrangian particles match the behaviour of fluid tracers in the simulations around the backflow regions. The ratio of jet width to length at late times is small, so we do not expect substantial effects of the increased jet width on large scale dynamics. The central panel of Figure~\ref{fig:method_summary} shows the density of jet material in the simulation (same region and slice as the top panel) together with the positions and velocities of a random subsample of the Lagrangian particles in this slice. 

\begin{table}
    \centering
    \begin{tabular}{c|c|c}
        \hline  Variable & Unit & cgs\\ \hline
         Length, $l_0$    & 1 kpc & $3.0856775807\times10^{21}$ cm \\
         Density, $\rho_0$ &  & $6.0\times10^{-27}$ g cm$^{-3}$\\
         Speed, $v_0$ & c & $2.99792458\times10^{10}$ cm s$^{-1}$\\ \hline
    \end{tabular}
    \caption{Simulation units}
    \label{tab:sim_units}
\end{table}

\subsection{Jet power timeseries}
\label{sec:jet_powers}

\begin{table}
    \centering
    \begin{tabular}{c|c|c}
        \hline
         $\sigma$ & Mean power over simulation (erg s$^{-1}$) & Seeds \\ \hline
         0.00    & $1.00\times 10^{45}$ & -  \\
         0.00    & $2.00\times 10^{45}$ & -  \\
         0.33    & $0.960\times 10^{45}$  & 0 \\
         0.33    & $0.815\times 10^{45}$  & 6 \\
         0.33    & $1.01\times 10^{45}$  & 12 \\ \hline
    \end{tabular}
    \caption{Simulation variables}
    \label{tab:sim_vars}
\end{table}
We aim to investigate morphological signatures of variability according to the predictions of CCA~\citep[][see Section~\ref{sec:intro}]{Gaspari2017RainingModel}. For this purpose, we simulate two jets with constant powers and three with variable power as detailed in Table~\ref{tab:sim_vars}. Here we motivate and describe the jet power distributions used in our variable jets.

In simulations of CCA~\citep{Gaspari2017RainingModel}, turbulence injected at large scales ($>4$ kpc from the black hole) leads to  thermodynamic fluctuations on smaller scales closer to the black hole. These fluctuations drive an accretion rate with a pink noise power spectrum and 
a lognormal distribution. $\sigma$ is the standard deviation of the natural logarithm of the accretion rate, which~\cite{Gaspari2017RainingModel} measure to be $\sigma=0.33$. For the low spin case, when $a\ll M$, the Blandford-Znajek mechanism predicts a jet power $Q$~\citep{Blandford1977ElectromagneticHoles}
\begin{equation}
    Q\propto0.3\left(\frac{a}{M}\right)^2L_D\propto10^{38}\left(\frac{a}{M}\right)^2\dot{M} \,\rm{W},
    \label{eqn:jet_power_scaling}
\end{equation}
where $a$ and $M$ are the black hole specific angular momentum in units of $c$ and mass in units of $c^2/G$, $L_D$ is the power radiation by the disc and $\dot{M}$ is the accretion rate in units of $M_\odot/\rm{yr}$. Furthermore, the linear dependence on accretion rate has been seen in GRMHD simulations of systems with high spin, with a spin-dependent efficiency factor~\citep{McKinney2005TotalDisk}. Assuming both CCA and the BZ mechanism are acting leads to the prediction of a lognormal distribution of jet powers (again with $\sigma=0.33$), with a pink noise power spectrum. Thus we set our variable jets to have powers obeying a lognormal probability density function
\begin{equation}
    p(Q) = \frac{1}{\sqrt{2\pi}\sigma Q}\exp\left[-\frac{(\ln Q - \ln Q_0)^2}{2\sigma^2}\right]\, ,
    \label{eqn:jet_power_pdf}
\end{equation}
where $\sigma=0.33$ is the standard deviation of the natural logarithm of the power and represents the variability of the source. $Q$ is the power at a given time and $Q_0$ is the median jet power. 

To simulate variable jets following this behaviour, we use an input time series which gives the speed of the jet at each time. We create these time series as follows. We first generate a time series for the jet power in our variable jets using the DELightcurveSimulation~\citep{Connolly2016DELightcurveSimulation:Code} implementation of the Emmanoulopoulos light curve simulation algorithm~\citep{Emmanoulopoulos2013GeneratingUpdated}. We specify a pink noise spectrum with power spectral density $S(f) \propto f^{-1}$ where $f$ is the frequency of the fluctuations and set $\sigma=0.33$.

\begin{figure}
    \centering
    \includegraphics[width=\linewidth]{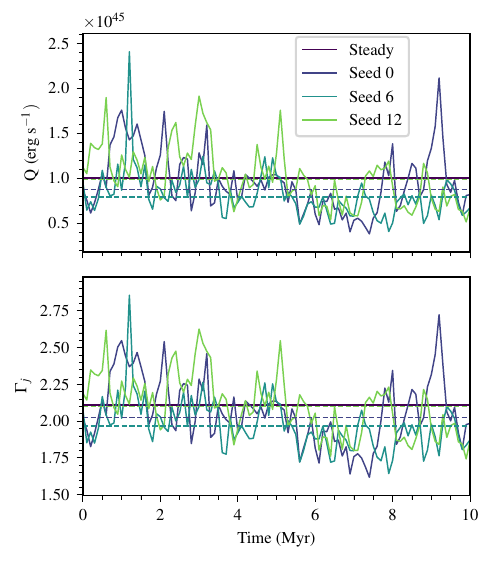}
    \caption{Input jet powers and corresponding Lorentz factors over time, with median jet powers and Lorentz factors over the 10 Myr sample shown by dashed lines (see also Table~\ref{tab:sim_vars}). There is an approximately linear scaling between $\Gamma$ and $Q$ when $\Gamma\approx2$.}
    \label{fig:jet_powers}
\end{figure}

The maximum age our simulations reach is limited by computational resource. As we are modelling jets which would continue to evolve beyond this maximum age, we generate time series 30 Myr long such that the median power of the jet over the 30 Myr is $10^{45}$ erg s$^{-1}$. We then use the first 10 Myr of the time series to evolve the simulations. This leads to differing median jet powers for the variable jets over the first 10 Myr as summarised in Figure~\ref{fig:jet_powers}. The jet power in RHD is given by~\cite[e.g.][]{Taub1948RelativisticEquations, Landau1987TheFields}
\begin{equation}
    Q=\pi r_j^2v_j\left[\Gamma_j(\Gamma_j-1)\rho_j c^2+\frac{\gamma_{\rm ad}}{\gamma_{\rm ad}-1}\Gamma_j^2P_j\right] \, ,
    \label{eqn:jet_power}
\end{equation}
where $\Gamma_j$, $v_j$, $\rho_j$ and $P_j$ are the jet Lorentz factor, speed, density and pressure at injection respectively, and $\gamma_{\rm ad}$ is the adiabatic index. Due to the supersonic nature of our jets, we follow~\cite{Whitehead2023StudyingJets} and neglect the energy contribution from the pressure when calculating $\Gamma_j$ from $Q$, and then numerically solve $Q=\pi r_j^2v_j\Gamma_j(\Gamma_j-1)\rho_j c^2$ for our adopted $\rho_j$ and $r_j$. The resulting Lorentz factors are shown in the lower panel of Figure~\ref{fig:jet_powers}. For our adopted radius, we obtain moderate Lorentz factors of $\approx2$, whereas for a decreased jet radius, Equation~\ref{eqn:jet_power} requires higher Lorentz factors to produce jets with the same power.

We enforce some conditions on the variability to ensure the stability of the simulations as sudden spikes in the jet Lorentz factor profile can cause numerical issues. We check that less than 60\% of the energy is injected into the domain in both the first and second halves of the time series, so as to avoid any time series which have very different amounts of energy injected in the first and second halves of the simulations. We use the first three seeds fulfilling this criteria -- seeds 0, 6 and 12 -- and generate the time series with a resolution of 100 kyr. Whilst we would expect variability on smaller timescales to be present, we find that simulations run using time series generated at higher resolution suffer numerical issues such as regions with negative pressures and densities near the jet base, due to larger changes in jet power over smaller time periods. We describe the potential effects of including variability on smaller timescales in Section~\ref{sec:limitations}. We save the emissivity information every 20 kyr. Material travelling at the speed of light can travel $\approx6$ kpc in this time, which is a tenth of the length of our simulation volume.
In addition to the steady and variable jets, we run four smaller simulations with discrete step changes in jet power from $10^{45}$ erg s$^{-1}$ to $2\times 10^{45}$ erg s$^{-1}$. These step-change simulations aim to investigate a simpler case of variability to improve our understanding of the mechanisms behind the morphological features we identify in the variable simulations. They use the results of the lower power steady jet at 3, 5, 7 and 9 Myr as input and comprise of a single immediate change in jet power to the higher power at those times.

\subsection{Acceleration of particles at shocks}
\label{sec:shock_acceleration}

\begin{table}
    \centering
    \begin{tabular}{c|c|c}
    \hline
         Parameter & Symbol & Value \\ \hline
         Threshold gradient of thermal pressure & $\epsilon_{\rm sh}$ & 1 \\
         Non-thermal fraction energy & $f_E$ & 0.1\\
         Non-thermal fraction particles & $f_N$ & 1\\
         Jet Internal Mach Number & $\mathcal{M}$ & 100 \\ \hline
    \end{tabular}
    \caption{Simulation parameters which are constant for all runs}
    \label{tab:sim_params}
\end{table}
The Lagrangian particle module~\citep{Vaidya2018AFlows} allows us to trace an electron energy spectra associated with each Lagrangian particle in the simulation. Here we describe the updates applied to those spectra as the Lagrangian particles pass through shocks in the simulation.

We identify shocks in the simulation as in~\cite{Vaidya2018AFlows}, identifying regions where the divergence of the velocity is negative and the difference in thermal pressure between cells obeys $\epsilon_{\rm sh} \equiv \left|\frac{\Delta P}{P}\right|> 1$. We use the convolution based approach to resetting the electron energy spectra at shocks as detailed by~\cite{Mukherjee2021SimulatingElectrons}. At shocks the electron energy spectrum is set to a power law $\mathcal{N}\propto E^{-q+2}$. As we do not have information on the geometry of the magnetic field (due to our choice to use RHD simulations), we cannot measure the angle between the shock and the magnetic field. We therefore use the perpendicular shock limit where the power law index is set to
\begin{equation}
    q=q_{\rm NR}+\frac{9}{20}\frac{r+1}{r(r-1)}\eta^2\beta_1'^2\, ,
\end{equation}
where $r$ is the compression ratio of the shock and $q_{\rm NR}$ is the limit for non-relativistic shocks, given by~\citep{Drury1983AnPlasmas}
\begin{equation}
    q_{\rm NR} = \frac{3r}{r-1}\, ,
\end{equation}
We adopt the perpendicular shock limit, rather than the parallel one for the following reasons. We are particularly interested in particle acceleration at the jet head.~\cite{Lopez-Miralles20223DBinaries} show that field lines around the head of a jet are quasi-perpendicular to the jet propagation direction in 3D RMHD simulations of jet-wind interactions in high-mass X-ray binaries. Observationally, radio polarisation maps have shown a magnetic field perpendicular to the direction of the jet~\citep{Godfrey2013AgnLobes}. In addition, ultra-relativistic shocks are characteristically quasi-perpendicular~\citep[e.g.][]{Kirk1999ParticleShocks,Sironi2015RelativisticMagnetization,Bell2018Cosmic-rayEstimates}. 
For strong relativistic shocks with $r>4$ the electron energy power law slope is set to $q=4.23$. 

The integrated energy transferred to non-thermal particles is set by a constant fraction $f_E$ of the energy density of the fluid~\citep{Mukherjee2021SimulatingElectrons}, $\rho e$, where we use $f_E=0.1$ throughout this work. This value reflects results from Particle in Cell (PIC) simulations~\citep[for a review see ][]{2020MarcowithMulti-scaleSystems}. Changes to $f_E$ would affect only the amount of energy given to non-thermal particles and not the slope of the injected distribution. Furthermore, the amount of energy injected does not affect the relative cooling rates between high and low energy electrons, so should not translate to significant changes in morphology. We would expect a higher value of $f_E$ to lead to a jet with an overall greater luminosity. There is an additional stipulation that there must be enough particles available to be shocked, accounting for electrons that have been accelerated at previous shocks. $f_N$ gives the fraction of the mass in the jet that is assumed to be electrons which can be accelerated at shocks. We set $f_N=1$, assuming that the jet is made only of electrons, although our results are unlikely to be significantly changed by assuming a different mass fraction. $\rho e$ is found by inverting the Taub-Matthews equation of state~\citep{Taub1948RelativisticEquations}~\citep[see Equations 8 and 13 of][]{Mignone2007EquationIndex}
\begin{equation}
    \rho e = \frac{1}{2}\left(3p - 2\rho c^2+\sqrt{(3p-2\rho c^2)^2+12p\rho c^2}\right)\, .
\end{equation}
\label{eqn:internal_energy}
The maximum energy that the non-thermal particles can reach is set by balancing the acceleration timescale with the synchrotron timescale~\citep{Bottcher2010TimingBlazars}. This gives a maximum energy, $\epsilon\propto\lambda_{\rm{eff}}^{-1/2}$, where $\lambda_{\rm{eff}}$ is the acceleration efficiency~\citep{Vaidya2018AFlows}. $\lambda_{\rm{eff}}$ is calculated using the quasi-perpendicular limit given in~\cite{Vaidya2018AFlows}. We estimate the required magnetic field strengths using $\rho e$. The magnetic field strength, $B$, is related to the energy density in the magnetic field, $u_B$, by
\begin{equation}
    B = \sqrt{2u_B}\, .
    \label{eqn:estimate_B_field}
\end{equation}
We assume that a constant fraction $f_B = \frac{u_B}{\rho e} = 0.1$ of the internal energy density is contained in the magnetic field, which given our choice of $f_E$ is equivalent to equipartition.

\subsection{Adiabatic cooling using a fluid tracer}
\label{sec:tracer_method_derivation}

\begin{figure}
    \centering
    \includegraphics[width=\linewidth]{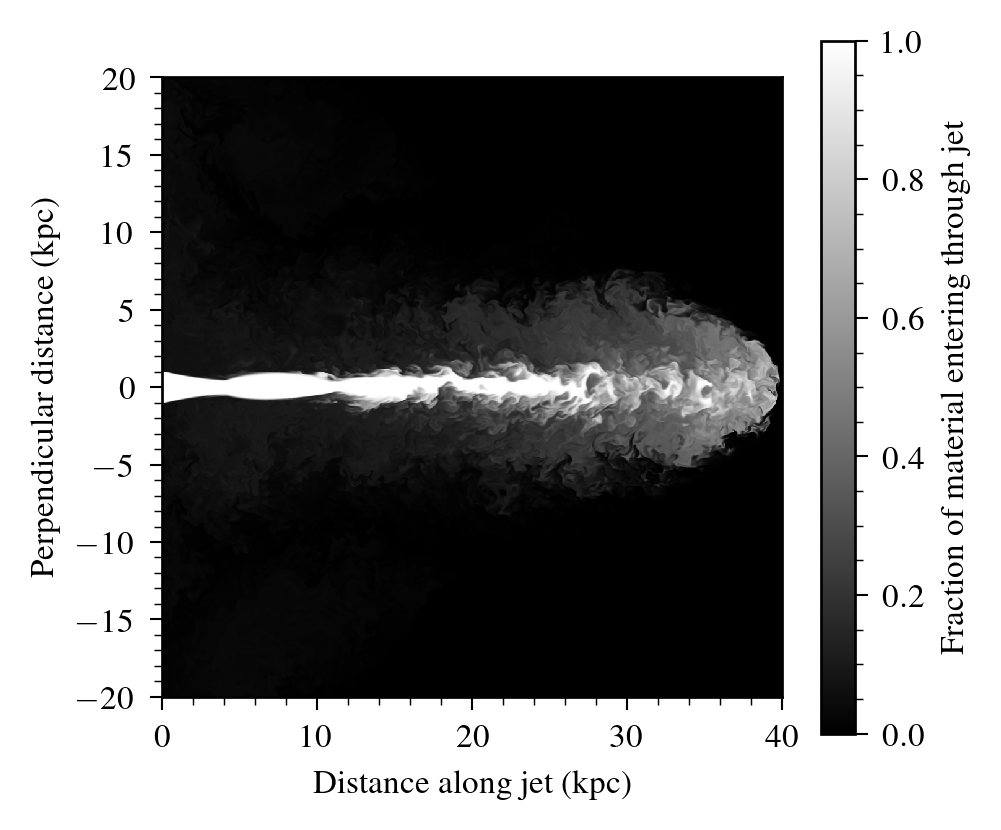}
    \caption{Passive scalar (fluid) tracer of jet material from central slice through the $10^{45}$ erg s$^{-1}$ constant power simulation at 7 Myr. Mixing of the low density material which enters the simulation through the jet with high density material originating in the ambient medium leads to very low fractions of jet material in the lobe regions.}
    \label{fig:tracer_example}
\end{figure}
As electrons move through regions of various densities, particularly from the jet head back into the lobes, they expand (and contract) leading to adiabatic cooling (and heating). The Lagrangian particle method~\citep{Vaidya2018AFlows} allows us to track the effects of this adiabatic expansion on the electron energy spectra of the electron population modelled by each Lagrangian particle.

We find significant mixing of jet material with other much denser material that originates in the ambient medium (see Figure~\ref{fig:tracer_example}). We treat the material which has passed through the bow shock as unshocked for the purposes of tracking the electron energy spectra. This is because the bow shock is not thought to be a site of efficient particle acceleration~\citep{2019MatthewsUltrahighGalaxies} compared to the hotspot in powerful radio sources, as it is not visible in well-resolved radio images of jets on scales of 10-100 kpc~\citep[e.g.][]{Perley1984TheA,Sabater2021TheField}. The inclusion of material accelerated at a bow shock could contribute to a faint diffuse radio component surrounding the main radio lobes, but this choice does not affect our main results and conclusions. To account for the mixing between these two types of material, we use the methods of~\cite{Vaidya2018AFlows} to evolve the energy spectra according to the adiabatic cooling, adapting them to include a factor of fluid tracer which measures the fraction of the density which originates in the jet. In the following derivation we assume that the tracer follows the material originating in a given shock perfectly. This would be the ideal approach, but to implement it would require a very large number of fluid tracers to be evolved, making it computationally unfeasible. We discuss the effects of instead using a tracer of jet material after the derivation.

Neglecting radiative losses, the evolution of the number density of non-thermal electrons per unit energy, $\mathcal{N}(E)$, is given by~\citep{Vaidya2018AFlows}
\begin{equation}
    \frac{d\mathcal{N}}{d\tau}+ \frac{\partial}{\partial E}\left(-\frac{E}{3}(\nabla_\mu u^\mu)\mathcal{N}\right) = -\mathcal{N}\nabla_\mu u^\mu\, ,
\label{eqn:N_evolution}
\end{equation}
where $\tau$ is the proper time in the comoving frame of the fluid and $u^\mu$ is the four-velocity of the fluid. We define $\chi = \mathcal{N}/(\mathcal{C}n)$, where $\mathcal{C}$ is a passive scalar advected with the fluid which acts as a tracer of shocked fluid (denoted $Q_k$ in the PLUTO documentation) and $n$ is the number density of all particles in the fluid. $\chi$ therefore gives the number of non-thermal electrons, but normalised to the number density of particles in the fluid that originated in the jet. Equation~\ref{eqn:N_evolution} then becomes
\begin{equation}
    \frac{d(\mathcal{C}n\chi)}{d\tau}+ \frac{\partial}{\partial E}\left(-\frac{E}{3}(\nabla_\mu u^\mu)(\mathcal{C}n\chi)\right) = -(\mathcal{C}n\chi)\nabla_\mu u^\mu\, .
    \label{eqn:Q_knchi_evolution}
\end{equation}
The fluid continuity equation for the shocked material can be written as
\begin{equation}
    \frac{d}{d\tau}(\mathcal{C}n)= - \mathcal{C}n\nabla_\mu u^\mu\, .
    \label{eqn:tracer_continuity}
\end{equation}
Combining equations~\ref{eqn:Q_knchi_evolution} and~\ref{eqn:tracer_continuity} gives the same form as in~\cite{Vaidya2018AFlows}:
\begin{equation}
    \frac{d\chi}{d\tau} + \frac{\partial}{\partial E}\left(-\frac{E}{3}(\nabla_\mu u^\mu)\chi\right) = 0\, ,
    \label{eqn:chi_differential}
\end{equation}
only now with the redefinition of $\chi$ as described above. Following the steps given by~\cite{Vaidya2018AFlows} gives
\begin{equation}
    \chi(E,\tau)dE=\chi_{0}dE_0\, ,
    \label{eqn:number_in_bin_conserved}
\end{equation}
where $\chi_{0}=\chi(E_0,\tau_0)$, which is a statement that the number of particles in an energy bin is conserved as the bin edges are moved according to adiabatic cooling or heating as described by 
\begin{equation}
    E(\tau) = E_0e^{-\int^\tau_{\tau_0}\frac{\nabla_\mu u^\mu}{3}d\tau}\, .
    \label{eqn:shrinking_spreading}
\end{equation}

Using Equation~\ref{eqn:tracer_continuity} to integrate the exponent gives
\begin{equation}
    E(\tau)=E_0\left(\frac{\mathcal{C}n}{\mathcal{C}_0n_0}\right)^\frac{1}{3}\, ,
    \label{eqn:bin_edge_evolution}
\end{equation}
which along with the condition of conserving the number of particles in a bin (Equation~\ref{eqn:number_in_bin_conserved}) gives
\begin{equation}   \chi(E(\tau),\tau)=\chi_0\left(\frac{\mathcal{C}(\tau)n(\tau)}{\mathcal{C}_0n_0}\right)^{-\frac{1}{3}}\, .
    \label{eqn:chi_update_tracer}
\end{equation}
The method described here so far uses a passive scalar tracer of shocked material, and assumes that this shocked material does not mix with other shocked material. For computational reasons (related to both cost and complexity) we cannot inject a distinct fluid tracer for every shocked region at every time. We instead make use of the jet tracer with some adaptations to the above method. Analysis of the tracer particles shows us that material injected late in the simulation mixes with material injected early in the simulation in the lobes of the jet. Using the jet tracer with the above method as described would lead to spurious adiabatic heating as recently injected and shocked material mixed into older shocked material in the lobes. We therefore choose to deactivate adiabatic heating in our scheme. To do this we enforce a check at every timestep on the density of jet material. If this quantity increases, we keep the electron energy spectrum constant, rather than allowing adiabatic heating. Through this, we assume that if a tracer particle moves into a region of higher jet material density this is due to mixing with previously shocked material, rather than due to compression of the recently shocked material represented by the Lagrangian particle. Along the jet, we expect this to have very little effect because whilst we might expect some adiabatic heating as electrons approach each shock along the jet, the energy spectra of the electrons is reset at each shock and this new addition of energy will dominate the emission in these regions. Furthermore, the jet tracer maintains a value of $\gtrsim0.5$ up to the hotspot of the jet (see Figure~\ref{fig:tracer_example}), such that the effects of our updated scheme on this region are small. 

Another effect of using the jet tracer rather than more direct tracers of shocked material is that we could underestimate emission from material that did not originate in the jet. On the one hand, this could underestimate emission from material which has been entrained along the jet line before being shocked in the hotspot, thus reducing both the hotspot luminosity and diffuse emission in the lobes. The aforementioned relatively high ($\sim0.5$) tracer values at the hotspot show that this is likely to be a small effect. On the other hand, the contribution of ambient medium material which has mixed with jet material in the backflow regions will be underestimated, with the effect of reducing the diffuse lobe emission, and the lower fraction of tracer in these regions suggests this latter effect may be more significant.

The emissivity calculation includes a factor of the density of jet material. This could lead to some overestimation of the amount of brightness of the emission from further back in the lobes due to the inclusion of jet material that has not been shocked or was shocked longer ago. However, we find that the visible regions in our log-scaled emissivity maps (see e.g. Figure~\ref{fig:emissivities_comparison}) do not include significant contributions from the lobes and as such any overestimation in these regions has a negligible effect on any conclusions that we draw.

Overall, our updated scheme improves predictions of lobe emissivity by removing spurious adiabatic heating in the case of strong mixing in a high density contrast simulation. We describe specific tests of the original and updated schemes in Appendix~\ref{sec:app_adiabatic_cooling}. It is likely that the neglect of adiabatic heating leads to overestimates of the amount of net adiabatic cooling, which then affects our predictions for the diffuse emission from the lobes. We therefore focus in particular on results relating to emission from along the jets and at the hotspots.

\subsection{Synchrotron cooling and emission from particles in RHD}
\label{sec:synchrotron_cooling}

We evolve the energy spectra of each electron population subject to synchrotron cooling as the Lagrangian particles move through regions of higher and lower magnetic fields. Similarly, the cooling due to inverse Compton scattering off the cosmic microwave background at redshift $z=0$ is also included with an effect equivalent to synchrotron cooling from an additional weak background magnetic field that is constant in space throughout the simulation domain. By evolving the spectra for each individual sample of electrons represented by each Lagrangian particle we account for the acceleration and cooling histories of the electrons informed by the properties of the jet that each population has experienced. This approach accounts for the different rates of synchrotron cooling in different areas of the jet, in addition to the different rates of cooling experienced by electrons at various energies.

To calculate the synchrotron cooling rate we require an estimate of the magnetic field strength at each point in the simulation. We do this as above using Equation~\ref{eqn:estimate_B_field}. In order to calculate the emission from each particle using our RHD simulations, we assume that the magnetic field is randomly distributed in the co-moving frame of each Lagrangian particle. We further assume that the electrons represented by the Lagrangian particle have an isotropic distribution of velocities in the co-moving frame (the limit of the isotropy assumption made in~\cite{Webb1989THEFLOWS}, in which small perturbations from an isotropic distribution are considered). We therefore assume isotropic emission in this frame, which is then Doppler-boosted for each Lagrangian particle individually in order to calculate the final emissivity along the line of sight to the observer. The lower panel of Figure~\ref{fig:method_summary} shows an example of an emissivity map created using this method. To create all emissivity maps shown in this work, we save the emissivities from the simulation on a per particle basis, deposit the emissivities to a grid with the same resolution as the underlying simulations and then apply a Gaussian filter. In this example image we use a standard deviation of 1 cell (0.04 kpc in each direction) for the Gaussian image, to highlight the similarities in fine structure with the pressure and density maps. In other images in this work we use a standard deviation of 2 cells (0.08 kpc in each direction). For reference, the 0.3 arcsecond resolution of LOFAR with international baselines would be able to resolve structures greater than 0.3 kpc large in a radio galaxy at a redshift of 0.05, corresponding to an area of approximately $8\times8$ cells in our simulations.

\section{Results}
\label{sec:results}

The modelling of energy spectra and emission in partnership with the passage of material through the jet using the Lagrangian particle method leads to a considerable improvement in the accuracy with which radio emission from these systems can be predicted. Furthermore, our novel approach to modelling adiabatic cooling using fluid tracer and Lagrangian particles in tandem greatly improves the modelling of mixing between non-thermal electron populations with other material. As a result we present radio emissivity maps showing key features found in radio galaxy jets (see Figures~\ref{fig:method_summary} and~\ref{fig:emissivities_comparison}). Across the jets we simulate we recover a range of morphologies, with changes in the degree of `edge brightening' and hotspot prevalence, with a variety somewhat reflective of that found in observations. We begin by examining the basic structure of the jets before moving onto discussion of the fiducial $10^{45}$ erg s$^{-1}$ constant power jet (Section~\ref{sec:fiducial}) and variable jets (Section~\ref{sec:variable_jet_results}) where we find large amounts of variability in the hotspot luminosity. To further our understanding of the mechanism driving this variability we examine the results of a second constant power jet with twice the power (Section~\ref{sec:higher_power}) and jets with step changes in power (Section~\ref{sec:step_change}). 

\subsection{Emissivity maps and shock features}
\label{sec:emissivity_maps}
The emissivity map shown in Figure~\ref{fig:method_summary} is an example from the variable jet with seed 12 at 9 Myr and shows some common features seen across our simulations (both constant power and variable), namely recollimation shocks, followed by a turbulent region, and finally a hotspot region at the head of the jet. Shortly downstream of the jet inlet, we see two prominent recollimation shocks, with a possible third at approximately 17 kpc. The exact luminosities and displacements of these regions relative to the jet base have been found to depend heavily on factors relating to the jet launch angle, speed and environment~\citep{Costa2025HowJets}, including the magnetic field structure at the base of the jet~\citep{Matsumoto2021MagneticJets}. We do not model the jet launch and acceleration regions here and so the scales and positions of the recollimation shocks in our simulations are instead a consequence of our jet injection prescription. We would expect recollimation shocks on parsec scales in real sources to provide a population of pre-accelerated electrons to the jet spine and jet head and the recollimation shocks in the simulated jets fulfil this role, contributing to the emission along the length of the jet. Strong emission along the jet line is seen in a variety of spatially well-resolved sources including Cygnus A ~\citep{Perley1984TheA}, Hercules A~\citep{Timmerman2022OriginObservations} and M87~\citep{Pasetto2021ReadingField}, and in many images of giant radio galaxies~\citep[e.g.][]{Simonte2024GiantFields}. Pictor A shows strong emission along the jet in X-rays, with knots which are also observed in radio and optical images~\citep{2016HardcastleDeepPictorA,2024AndatiSpectropolarimetricMeerKAT,2015GentryOpticalJet}. If this emission is caused by synchrotron emission, it necessitates in-situ particle acceleration on kpc scales~\citep{2016HardcastleDeepPictorA}. The approximately regular spacing and number of these knots in the radio and X-ray images is reminiscent of the recollimation shock structures in our simulations. Further to this, because hydrodynamic simulations are scale invariant, our recollimation shocks highlight interesting phenomena relating to the interplay between variability and recollimation shock movement (see Section~\ref{sec:variable_jet_results}). 

In agreement with previous work~\citep{Massaglia2016MakingJets,2018GourgouliatosReconfinementNuclei,Costa2025HowJets, 2025BoulaInstabilitiesJets}, the recollimation shocks in Figure~\ref{fig:method_summary} begin to be disrupted by instabilities at larger distances, leading to patchy emission spread over a larger cross sectional area. Despite these instabilities, our simulations retain a hotspot throughout the evolution of the steady jet and for much of the time in the variable jet cases (see Figure~\ref{fig:emissivities_comparison}), most likely due to the relatively high jet power used in our simulations. The turbulent region is visible in the emissivity maps as a collection of small cloud-like structures, which often appear elongated along oblique angles to the jet direction. The high resolutions used in these simulations allow a particularly detailed view of this region, revealing structures that would otherwise not be seen. Some caution should be applied when comparing the emissivity map -- which is integrated along the line of sight -- with the density and pressure images, which are slices through the domain. Nonetheless, Figure~\ref{fig:method_summary} shows that these elongated, brighter structures are associated with regions of high pressure, which often have sharp boundaries, indicating shocks. Moreover, the density slice highlights the turbulent mixing of recently injected jet material with material filling the cocoon region.

A hotspot forms at the head of the jet, associated with a strong termination shock. This hotspot corresponds to a region of particularly high pressure. As can be seen in Figure~\ref{fig:tracer_example}, this area is still made up of $\gtrsim50\%$ jet material. In Figure~\ref{fig:method_summary}, the hotspot is spread out over a volume of a few cubic kpc, but contains a particularly bright spot just below the central jet axis. A region of backwards flowing material can be seen around the jet head in the emissivity maps, although the extent of this region is relatively modest, extending back no more than a few kpc in the direction of the jet axis. The Lagrangian particles shown in Figure~\ref{fig:method_summary} show that the hydrodynamic backflows extend much further than those visible in the emissivity maps.

\subsection{Fiducial constant power \texorpdfstring{$\mathbf{10^{45}}$}{1e45} erg\texorpdfstring{ $\mathbf{\rm{s}^{-1}}$}{/s} jet}
\label{sec:fiducial}

\begin{figure*}
    \centering
    \includegraphics[width=0.95\linewidth]{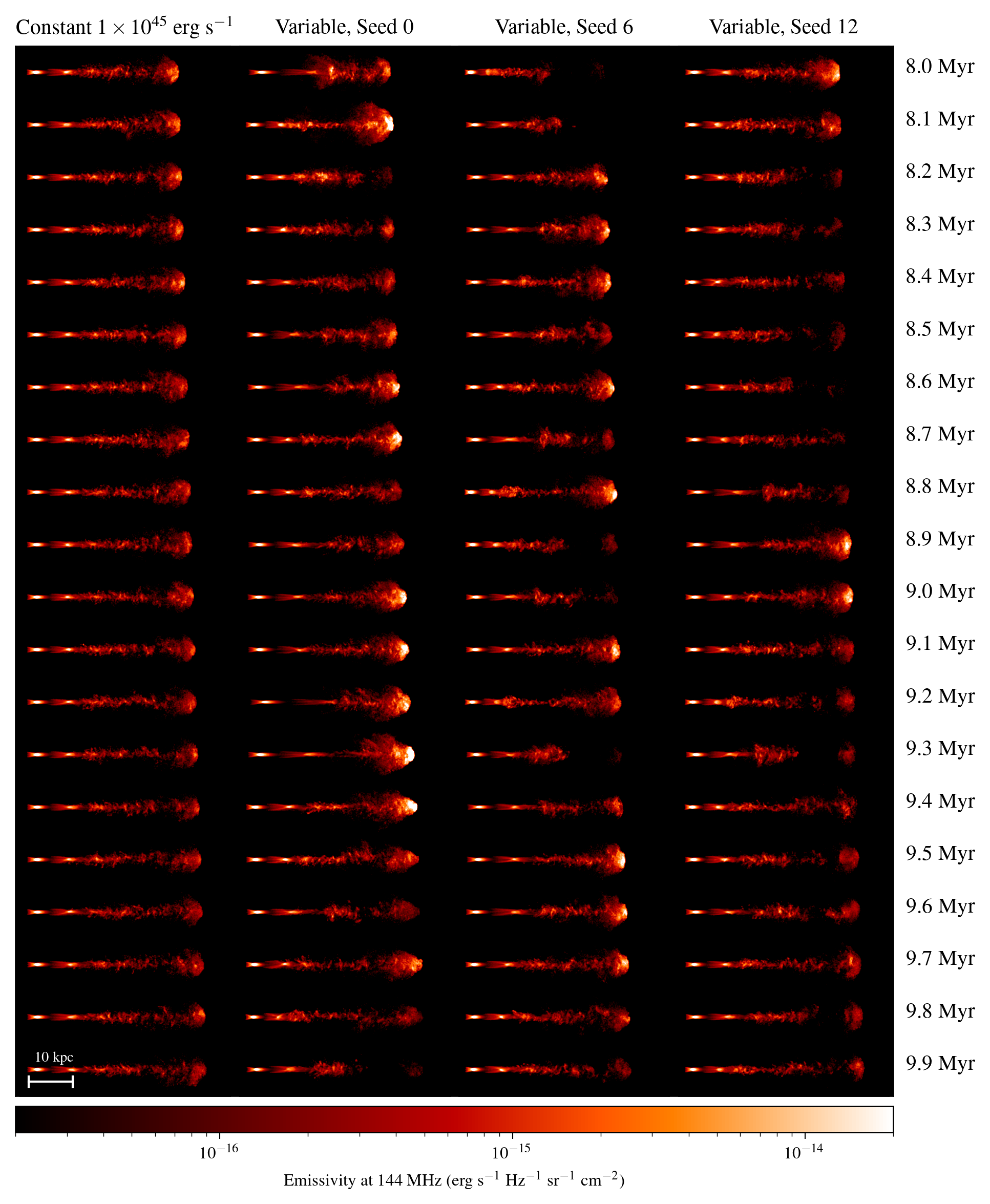}
    \caption{Emissivity maps for the 8-10 Myr period of the constant power $10^{45}$ erg s$^{-1}$ jet and the three variable power jets with seeds 0, 6 and 12. The constant power jet has a similar morphology over this full time period with two to three clear recollimation shocks which transition into a turbulent region followed by a modest hotspot. In contrast, the variable jets show a large amount of variety in their morphologies over time. The recollimation shocks move to larger and smaller distances from the jet launch point and the hotspot luminosity changes dramatically over the 2 Myr, sometimes being much brighter than the hotspots in the constant power case. See supplementary material for videos of the variable jets (\href{https://youtu.be/cen9ZF1PPRM}{seed 0}, \href{https://youtu.be/mk9wSW6_E_Q}{seed 6}, \href{https://youtu.be/113B7f87fRA}{seed 12}).}
    \label{fig:emissivities_comparison}
\end{figure*}

\begin{figure}
    \centering
    \includegraphics[width=\linewidth]{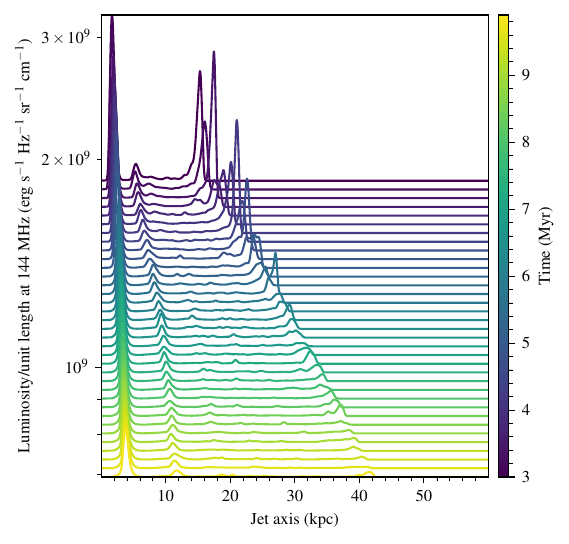}
    \caption{Luminosity at 144 MHz as a function of distance along the jet for the jet with constant power $10^{45}$ erg s$^{-1}$. Lines are offset from zero depending on jet age with earlier times in the simulation shown further up the axes. The emissivity maps are convolved with a Gaussian filter with a standard deviation of 5 prior to integration.}
    \label{fig:emission_vs_z}
\end{figure}

Whilst Figure~\ref{fig:method_summary} shows a specific moment in a variable jet's evolution as an example, the generic features of recollimation shocks, turbulence and hotspot can also be seen throughout the last 2 Myr of our fiducial $10^{45}$ erg s$^{-1}$ constant power jet, shown in the first column of Figure~\ref{fig:emissivities_comparison}. Although we only include the last 2 Myr in this image, we point the reader to animations of the evolution of each of the jets included in this over the 4-10 Myr period. Figure 4 shows a systematic decrease in the luminosity of the constant power jet over time with some slight variation in the luminosity of the hotspot. In Figure~\ref{fig:emission_vs_z} we sum all emissivities over the directions perpendicular to the jet direction to leave the total luminosity at 144 MHz as a function of distance along the jet.

There are two clear recollimation shocks at all times, with a third sometimes visible. The recollimation shocks move to larger distances over time, but do so in a steady and very gradual way. This is in agreement with the predictions of \cite{1991FalleSelf-SimilarJets}, who give an expression for the position at which the first recollimation shock in a self-similar jet should occur as
\begin{equation}
    z = \theta\left(\frac{\mathcal{K}}{p_c}\right)^{\frac{1}{4}},
\end{equation}
where $z$ is the distance along the jet, $\theta$ is the half-opening angle, $p_c$ is the cavity pressure and $\mathcal{K}$ is given by
\begin{equation}
    \mathcal{K}=\frac{2}{\gamma_{\rm ad}+1}\frac{(2Q_j\dot{M}_j)^\frac{1}{2}}{\pi\theta^2},
\end{equation}
where $\dot{M}_j=\pi r_j^2\rho_jv_j$ is the mass flow rate through the jet base. For our jets this suggests a dependence of $z$ on $v_j$ given by
\begin{equation}
    z\propto\frac{v_j}{(1-v_j^2)^{\frac{1}{4}}}.
\end{equation}
This model predicts that the height of the recollimation shocks should increase with decreasing cavity pressure, albeit without accounting for relativistic effects. In a qualitative sense, then, the model is in agreement with the gradual movement of recollimation shocks in our constant jet power simulation. We discuss the movement of the recollimation shocks in our simulations due to changes in jet power further in Section~\ref{sec:variable_jet_results}. 

The luminosity of the jet decreases substantially over time (as predicted by models of radio galaxy evolution~\citep[e.g.][]{2003PeruchoDynamicalSources,2018HardcastleSimulationGalaxies,2025BeltranPalauActiveGalaxies} and in particular, the hotspot decreases in luminosity steadily as seen in the images of the last 2 Myr. The jet hotspot can be seen to advance more slowly at later times. The images (Figure~\ref{fig:emissivities_comparison}) do show time variation in the detailed structure of the turbulent regions of the jet and lobe; however on macroscopic scales, both the images (Figure~\ref{fig:emissivities_comparison}) and luminosity profiles (Figure~\ref{fig:emission_vs_z}) illustrate that the steady jet displays slowly changing, predictable behaviour without significant variations in morphology beyond a change in length.

\subsection{Flickering jets}
\label{sec:variable_jet_results}

In contrast to the constant power jets, Figure~\ref{fig:emissivities_comparison} shows that the morphologies of the variable power jets change significantly over the course of the last two Myr of the simulation. In some cases, there is a clear hotspot and the jet morphology resembles that of a classical FRII radio galaxy. However, in others, the jet is more disrupted, due to a drop in power, and the morphology is more reminiscent of an FRI-like source. The change between these two morphologies can occur relatively quickly, sometimes over $\sim 0.1$ Myr timescales.

The hotspots are perhaps the most obviously variable feature of the morphology. There are periods with hotspots significantly brighter than those of the steady jet in all three seeds during this 2 Myr period, with particularly bright hotspots in the case of seed 0. Equally, there are periods where the hotspot is either very faint or not visible at all in the variable cases. From Figure~\ref{fig:emissivities_comparison} we can already see that the hotspot variability dominates the variability in the luminosity of these sources.

\begin{figure*}
    \centering
    \includegraphics[width=\linewidth]{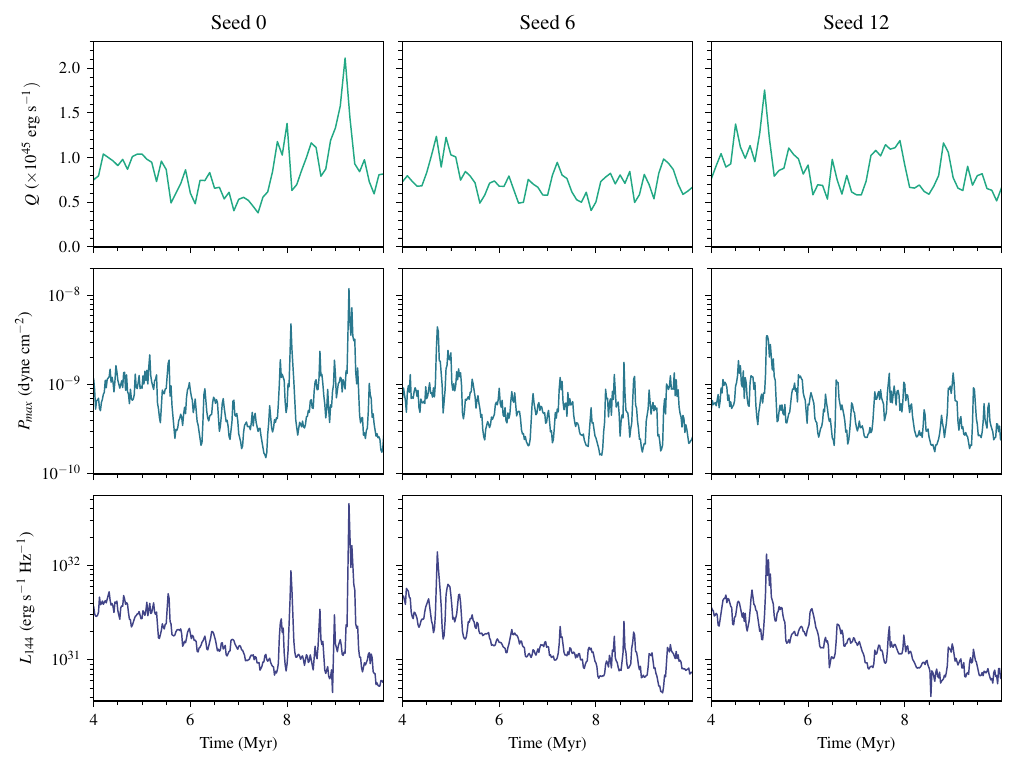}
    \caption{Jet power, maximum pressure at the hotspot ($P_{\rm max}$) and luminosity of the jet at 144 MHz over time for the three variable jet simulations from 4 to 10 Myr. A moderate increase in the jet power can cause a dramatic increase in the maximum pressure at the hotspot and in the luminosity of the full jet. The spikes in luminosity are closely related to the increase pressure at the hotspot, indicating that this region is of key importance in understanding the effects of variability. There is an underlying general trend towards decreasing luminosity over time which is also reflected in our constant power jet simulations (see Figure~\ref{fig:steady_jets}).}
    \label{fig:luminosity_vs_time}
\end{figure*}

With the aim of understanding the evolution of the behaviour of the variable jets over a longer time period, and of beginning to understand the mechanism behind the large variations in hotspot luminosity, we show the jet power, maximum pressure at the hotspot and luminosity of the source at 144 MHz against the jet age in Figure~\ref{fig:luminosity_vs_time}. To find the maximum pressure at the hotspot, we first subsample the pressure data, averaging the pressure across regions of $10\times10\times10$ cells. We then work backwards from the jet head and find the maximum pressure within 5 kpc along the jet axis. The hotspot pressure and luminosity for each seed show a combination of a systematic decrease with time (seen also in the constant jet case, see first column of Figure~\ref{fig:emissivities_comparison}) with the variable behaviour imprinted by the flickering jet power. There is a clear correlation between spikes in the jet power, and spikes in the pressure of the hotspot and the luminosity, with an increasing time lag as the jet increases in age and length due to the travel time of material along the jet. Whilst the increases in jet power are around a factor of 2 at most, the increases in pressure and luminosity can be of order 10 times or larger. We conclude that the dramatic increases in the overall luminosity of the jet are driven primarily by the increased hotspot pressure. We discuss the mechanism behind this in Section~\ref{sec:step_change}.

The behaviour of the recollimation shocks is also in significant contrast to the constant power jet case. Here the variability in jet power can be seen to lead to both increases and decreases in the distance along the jet at which the recollimation shocks occur, with these changes happening over short amounts of time (again of order $<0.1$ Myr). Indeed the long term behaviour of the recollimation shocks in a steady jet can be seen in Figure~\ref{fig:luminosity_vs_time}.

\begin{figure}
    \centering
    \includegraphics[width=\linewidth]{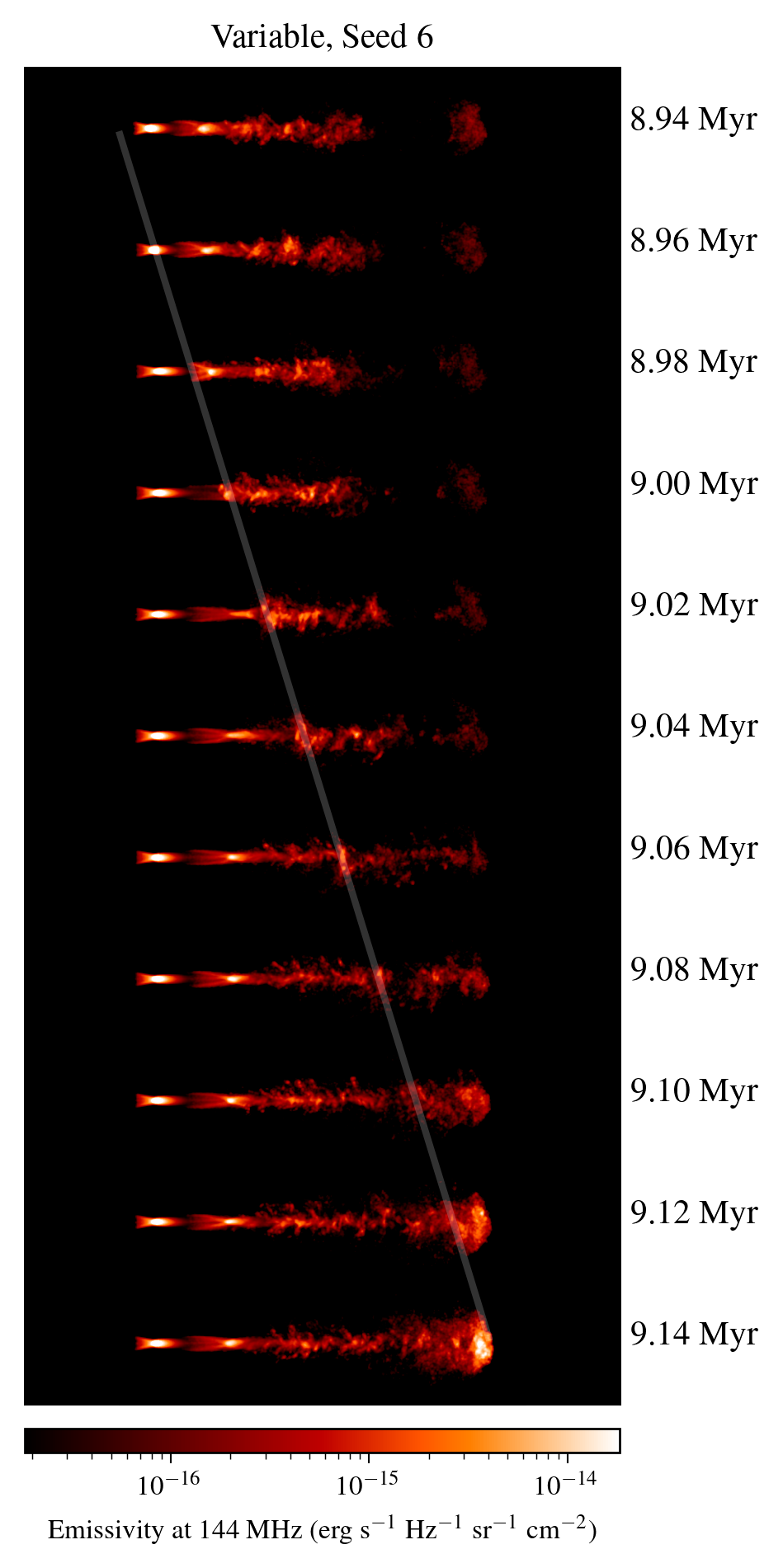}
    \caption{Variable jet with seed 6 between 8.94 and 9.14 Myr, showing the disruption and reformation of the recollimation shocks (8.98 to 9.02 Myr) and the passage of a bright region of emission along the jet (most clearly visible between 9.02 and 9.06 Myr). A line is overlaid to highlight the passage of the bright region along the jet between 9.02 and 9.06 Myr, and is continued backwards and forwards in time assuming a disruption propagating with a constant velocity. This highlights the coincidence with the disruption of the recollimation shocks and the brightening of the hotspot.}
    \label{fig:travelling_shock}
\end{figure}

We turn now to look more closely at the connection between the changes in recollimation shock separation and the brightening hotspots. There are times where the first two recollimation shocks are very close together, often with a suggestion of a third and sometimes even fourth shock visible. At other times the recollimation shocks are separated by larger distances. The movement of recollimation shocks to larger distances is predicted by~\cite{1991FalleSelf-SimilarJets} to happen at higher jet powers (see Section~\ref{sec:fiducial}). In our simulations this movement generally precedes snapshots with brighter hotspots, with a return to closer together recollimation shocks preceding dimming hotspots. To illustrate this Figure~\ref{fig:travelling_shock} shows the variable jet with seed 6 between 8.94 and 9.14 Myr at a finer time resolution than in Figure~\ref{fig:emissivities_comparison}. At 8.94 and 8.96 Myr the two recollimation shocks are close together. Between 8.98 and 9.00 Myr the second recollimation shock is disturbed -- it reforms at 9.02 Myr at a larger distance from the first recollimation shock. We use a faint overlaid line to show the position of a relatively bright patch of emission which appears to travel along the jet. This is most clearly visible between 9.02 and 9.06 Myr. We here direct the reader to the videos of the varying jets provided in the supplementary material (\href{https://youtu.be/cen9ZF1PPRM}{seed 0}, \href{https://youtu.be/mk9wSW6_E_Q}{seed 6}, \href{https://youtu.be/113B7f87fRA}{seed 12}), where further examples of this behaviour can be seen. In Figure~\ref{fig:travelling_shock} we continue the faint line backwards tracing the presumed path of this bright patch back in time and back towards the jet inlet, assuming it has a constant speed along the jet. The timing lines up with the disturbance of the second recollimation shock at 8.98 Myr. Assuming the bright patch is related to a disturbance that continues to propagate forwards in time at this constant speed along the jet, even past the point at which it stops radiating enough energy to be bright in the simulated radio image, leads to it reaching the position of the jet head at the same time as which the jet head brightens. Figure~\ref{fig:luminosity_vs_time} shows that there is a modest but sharp increase in jet power just before 9.0 Myr. This is followed by a short period of slower increase in power around 9.0 Myr and then shortly followed by another sharp increase. In summary, we observe that high power periods lead to an increase in the separation between the recollimation shocks, followed by a transient period of dramatically increased hotspot luminosity.

\subsection{Effects of a constant larger power}
\label{sec:higher_power}

\begin{figure}
    \centering
    \includegraphics[width=\linewidth]{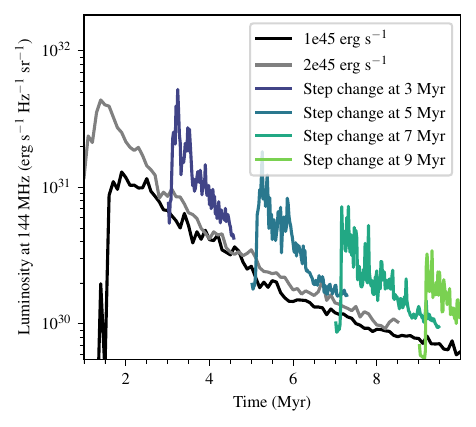}
    \caption{Luminosity against time for the steady and step-change jets. For each of the step-change jets, the luminosity grows sharply to a maximum a short time after the step change in power is introduced. The increased luminosity reduces to be comparable to that of the constant power jets over a period of $\Delta t\approx2$ Myr. The reduction in luminosity is not smooth, suggesting some re-brightening in cases. The initial jump in luminosity is strongly dependent on the age of the jet. Note the different sampling frequencies for the steady and step-change jets.}
    \label{fig:steady_jets}
\end{figure}

We have shown that dramatic increases in hotspot luminosity are associated with periods of high jet power and that in specific examples we can link these to earlier changes in the separation between recollimation shocks and regions of bright emission that travel along the jet. This suggests a link between the variability and the increased luminosity. Here we seek to understand whether the link to higher power periods is intrinsically linked to the variability or alternatively whether a constant larger jet power would be sufficient to drive the full extent of the increased luminosity. Observationally, \cite{Rawlings1991EvidenceSources} found the boundary between FRI and FRII galaxies to be at a jet power of approximately $10^{45}$ erg s$^{-1}$, however recent simulations have seen lower boundaries at approximately $10^{43}$ erg s$^{-1}$~\citep[e.g.][]{Massaglia2016MakingJets}, whilst~\cite{Mingo2019RevisitingLoTSS} have found FRII radio galaxies at low luminosities, with significant overlap between the distributions of luminosities for FRI and FRII sources. With jets above the FRII threshold power (which is not unique and, minimally, an environment-dependent quantity) expected to have bright hotspots and jets below this threshold expected not to show bright hotspots, we consider the possibility that our simulations are near to the threshold and that simply the higher power is what drives the dramatic increase in hotspot luminosity, rather than the variability itself. Therefore we consider a jet with a constant power of $2\times 10^{45}$ erg s$^{-1}$ -- twice that of our fiducial steady jet, and higher than all but the two largest spikes in power for our variable jets. Figure~\ref{fig:steady_jets} shows that the $2\times 10^{45}$ erg s$^{-1}$ jet is more luminous than the lower power steady jet at early times; however, at late times the luminosities of the two jets differ by less than a factor of 2 at the same timestamp. It should be noted that the $2\times 10^{45}$ erg s$^{-1}$ jet advances at a greater speed into the ambient medium, resulting in a jet that is longer for the same width at the base compared to the $10^{45}$ erg s$^{-1}$ constant power jet. We find similar late-time morphologies in the $2\times 10^{45}$ erg s$^{-1}$ and $10^{45}$ erg s$^{-1}$ cases in our simulation. We therefore find that a constant higher jet power is not sufficient to cause the very bright hotspots in our variable simulations. Rather, we conclude that this behaviour is directly linked to the {\em changes} in jet power and by extension that any variability in jet power is likely to lead to periods of significantly increased hotspot luminosity and a temporary enhancement to the jet radiative efficiency.

\subsection{What mechanism drives increases in hotspot luminosity?}
\label{sec:step_change}

\begin{figure*}
    \centering
    \includegraphics[width=0.95\linewidth]{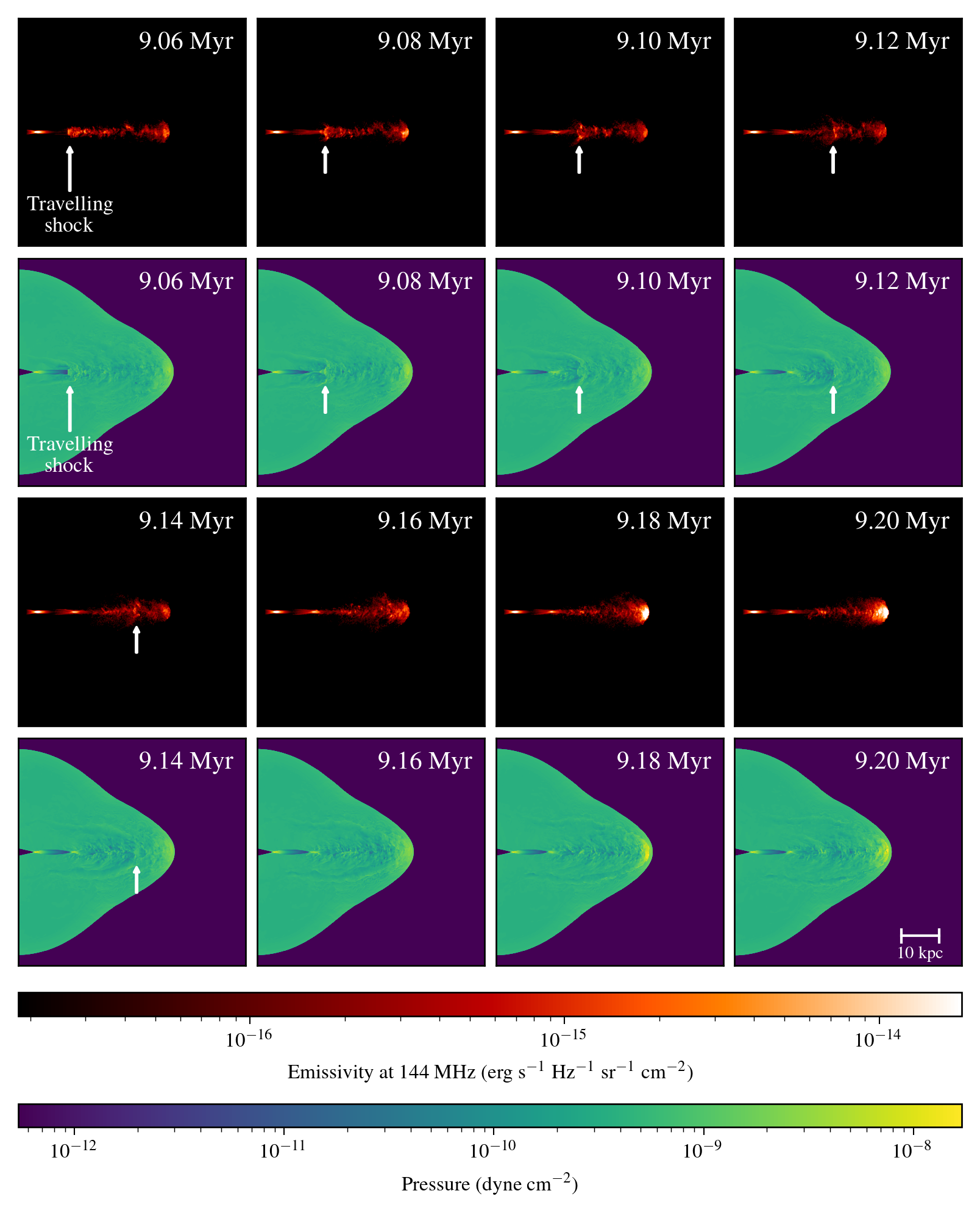}
    \caption{The effects of a step change in jet power from $10^{45}$ erg s$^{-1}$ to $2\times 10^{45}$ erg s$^{-1}$ at 9 Myr on emissivity at 144 MHz and pressure. The pressure is shown for a slice through the centre of the simulation domain (perpendicular to the line of sight shown in the emissivity maps). The box shown in each case is 60 kpc by 60 kpc. A shock can be seen moving along the jet axis in the pressure images, accompanied by a pressure wave moving outwards into the cocoon. This shock is also visible in the emissivity maps as a region of increased emissivity which is elongated perpendicular to the jet axis. The approximate displacement along the jet is denoted with an arrow, up until the travelling shock approaches the hotspot region. When the shock reaches the end of the jet there is a dramatic increase in hotspot pressure, corresponding to a large increase in the hotspot luminosity. We include videos of the step change emissivity maps and pressures in the supplementary material (\href{https://youtu.be/cpoXzzLl7zg}{144 MHz emissivity map}, \href{https://youtube.com/shorts/Btu8PY43m2w?feature=share}{pressure slice}).}
    \label{fig:step_change_combined}
\end{figure*}

To further investigate the mechanism driving these sharp increases in luminosity, we take the simpler case of a step change in jet power between the powers of the two steady jets previously discussed. We increase the power from $10^{45}$ erg s$^{-1}$ to $2\times 10^{45}$ erg s$^{-1}$, repeating this experiment with the step change occurring at 3, 5, 7 and 9 Myr. Figure~\ref{fig:steady_jets} shows that following a step change in power, there is a jump in luminosity by a factor of $\approx 8$ which gradually reduces down towards the luminosity of a constant power jet. The increase in luminosity caused by the change in jet power is strongly dependent on the age of the jet. This is likely to be caused by the reducing pressure of the cocoon over time in the steady jet simulations. This reducing pressure with increasing lobe length has been predicted using analytical estimates~\citep[e.g.][]{1989BegelmanOverpressuredSources}. We note times when the luminosity re-brightens, which are particularly clear in the step changes at 3 and 5 Myr. Overall, the sharp increases and the corresponding decays to steady jet behaviour clearly show that the variability indeed drives the increased luminosity, as opposed to purely the magnitude of the power at any given time.

The pressure and emissivity maps at times shortly following one of these step changes give insights into the process driving the jumps in luminosity. We show emissivity maps together with the pressure of a slice through the centre of the simulation domain for the step change at 9 Myr in Figure~\ref{fig:step_change_combined}. The formation of a clear hotspot is visible at 9.18 Myr and this hotspot can be seen to decay in luminosity from approximately 9.24 Myr. A shock travels along the jet axis of the simulation, accompanied by a pressure wave propagating into the cocoon region. From 9.06 - 9.14 Myr there is a bright region of emission coincident with the location of the travelling shock. In a similar manner to that highlighted for the jet shown in Figure~\ref{fig:travelling_shock}, this region is not visible in the emissivity map as the disturbance reaches the turbulent region (here at around 9.16 Myr). Additionally, at this time there is no clearly discernable shock wave along the jet axis in the pressure map. At 9.18 Myr the shock is once again identifiable in the pressure map and corresponds to the bright hotspot region in the emissivity map. We therefore identify the primary mechanism driving the increased luminosity following jumps to higher power periods in our variable jets as follows. AGN jet head regions consist of a forward and reverse shock separated by a contact discontinuity~\citep{Blandford1974ASources}. The sharp rise in power seen at many times during our simulations leads to a shock structure travelling along the axis of the jet, which, at a time approximately equal to the light travel time down the jet, will reach the existing double shock structure at the jet head and will interact with it. These interacting shocks lead to a large increase in the pressure in a thin region close to the jet head, which drives the increased luminosity. We plan to model the expected increase using semi-analytic methods in future work.

We expect the extent of the pressure jump resulting from this `interacting shocks' mechanism to also depend on the properties of the existing shock at the hotspot and on the effects of instabilities along the jet axis, which themselves may depend on the previous variability in jet power. This complexity further motivates the need to use high resolution simulations in the general case to understand the expected magnitude of the effects of variability.

In the context of compact jets in X-ray binaries, \cite{Malzac2014TheShocks} describe a forward and reverse shock structure at the boundary between two shells, where the forward and reverse shocks separate in the frame of the centre of momentum of the two shells. As shells collide, they become strongly compressed, leading to regions with increased pressure. In the methods employed in our simulations, this rise in pressure leads to increased emission in this region, as not only do we estimate higher magnetic fields in higher pressure regions, so that the non-thermal electrons in these regions lose energy at higher rates, but our prescription also transfers more energy to non-thermal particles at shocks in areas with higher internal energy. The travelling shocks identified in Figures~\ref{fig:travelling_shock} and~\ref{fig:step_change_combined}) represent the boundary between two `shells' in our simulation. On a microphysical scale, we might expect amplified magnetic field strengths due to the compression of shells, leading to higher synchrotron cooling rates and an increased efficiency of particle acceleration. The travelling shock shown in Figure~\ref{fig:step_change_combined} represents the boundary between two `shells' in our simulation. Due to the longer timescales of variability here, the travelling shock reaches the jet head before the shells merge.

\section{Discussion}
\label{sec:discussion}
\subsection{Implications for interpreting observations}

The results of our simulations have significant implications for interpreting observations of radio galaxies, which have a large variety of morphologies and luminosities. Differences in appearance and luminosity between radio galaxies are often attributed to long-term ($\gtrsim10$ Myr) differences between individual sources such as environment~\citep[e.g.][]{Saxton2010Time-dependentJets}, merger history~\citep[e.g.][]{RamosAlmeida2012AreInteractions} or age~\citep[e.g.][]{Shabala2008TheGalaxies}. Our results show that an individual source can exhibit very different structures and morphologies within small fractions of the source's overall lifetime. This view of variation within a single source both further establishes a relationship between the morphology of a source and its underlying fuelling mechanism and could go some way to providing a more unified view of processes happening across a large sample of radio galaxies. In the discussion that follows we consider some of the implications our results may have for interpreting observations of radio galaxies. 

\subsubsection{Luminosity-size evolution}
Models of tracks through luminosity-linear size (P-D) diagrams tend to use constant power jets to explain the distribution of sources across the plane. Our results support the observed trend for decreasing luminosity with size~\citep[e.g.][]{Blundell1999TheSamples, Shabala2008TheGalaxies} but suggest that some of the scatter in the distribution could be driven by variability in the jet power and resulting luminosity of single sources, rather than purely differences between the power (and environments) of various sources. We continue by looking at estimates of jet power, particularly in relation to radio galaxies with very bright hotspots. In these cases, we suggest that estimates of instantaneous jet power should consider the possibility that the hotspot luminosity is being largely driven by the merging of two sets of shocks caused by a recent sudden large increase in jet power. To make a speculative estimate of how often a significant luminosity increase occurs, we make use of the luminosity time series presented in the bottom row of Figure~\ref{fig:luminosity_vs_time}. We calculate the fractional difference in the luminosity at each time compared to its value 0.1 Myr earlier ($L(t)/L(t-0.1\rm\,{Myr})$). We then count the number of distinct times this value rises above $2$, $5$ and $10$ for each seed. We present the results of this analysis in Table~\ref{tab:luminosity_increases}. Whilst a larger sample of random seeds would clearly improve the ability to draw meaningful conclusions, particular for the more extreme increases, the values found begin to give us an idea of how often we expect to see significant increases, for example based on these results we might expect to see a luminosity increase of greater than $5\times$ once every approximately 4 Myr. The hotspot brightness remains at an elevated brightness compared to the baseline luminosity for a time on the order of 0.2 Myr, implying that we might expect approximately 1 in 20 sources to have some hotspot brightening due to this mechanism, with many of those sources showing low levels of brightened hotspots. Given the relatively quick decay in luminosity and the need to differentiate increases due to interacting shocks from increases due to environmental factors, we suggest that future work should focus on how often the more extreme increases are likely to occur. In relation to the link between morphology and fuelling we suggest that with a wider sample of well-resolved radio jets, key signatures found in our simulations together with measurements of core luminosities could contribute to furthering understandings of fuelling on timescales of fractions of a Myr.
\begin{table}
    \centering
    \begin{tabular}{c||c|c|c}
    \hline
         Luminosity Increase &Seed 0&  Seed 6 & Seed 12\\ \hline
         $>2\times$& 6& 7&6\\
         $>5\times$& 2& 1&2\\
         $>10\times$& 2&0 &1 \\ \hline
    \end{tabular}
    \caption{Number of times the luminosity increases by a given fraction between 4 and 10 Myr in each variable power jet simulation.}
    \label{tab:luminosity_increases}
\end{table}
A common method to investigate AGN jet evolution using population studies is to plot samples on a luminosity-linear size (P-D) plane and then compare to smooth tracks based on models with constant jet power~\citep[e.g.][]{Blundell1999TheSamples}.~\cite{Hardcastle2019Radio-loudSources} highlight that radio luminosity and size depend on the source environment, angle of a source to the line of sight and redshift in addition to the the power and age of the source. These attributes that vary between sources can be used, together with a population of sources with a variety of constant jet powers, to explain the large amounts of scatter and overlap between groups of sources at different redshifts~\citep[e.g.][]{Blundell1999TheSamples}. We consider whether the variable jets we simulate might also lie within the observational constraints. Tracks in luminosity-linear size diagrams are influenced by the advance speed of jets together with their luminosity. We consider the effects of each of these factors on the paths of our variable jets through the P-D plane.~\cite{Whitehead2023StudyingJets} show that, during high power periods, variable jets with $\sigma = 1.5$ (where $\sigma$ is the standard deviation of the logarithm of the accretion rate, see Section~\ref{sec:jet_powers}) have an increase in axial ratio $\mathcal{A}=\mathcal{L}/\mathcal{W}$ of the jet, where $\mathcal{L}$ is the jet length and $\mathcal{W}$ is the bow shock width measured at the jet base. This suggests that the force of the jet on the surrounding medium is concentrated over the area at the head of the jet during these times. Additionally \cite{Whitehead2023StudyingJets} show that these high power periods preferentially create hotspots due to larger instantaneous ram pressure at the jet head. In this work, with $\sigma=0.33$ as motivated by CCA (see Section~\ref{sec:jet_powers}), we instead find that the advance speeds of the variable jets we simulate vary only slightly from the advance speeds of our steady jets, with both advancing at a near constant speed. This suggests that the main effect of any variability resulting from CCA on predictions for luminosity-size diagrams is through the resulting variability in the luminosities. This is characterised by an overall trend towards decreasing luminosity towards larger linear sizes but with significant scatter introduced by the varying hotspot luminosity. The variable hotspot luminosity of our variable power jets could provide an alternative or additional explanation for this scatter, whilst still following the overall constraint of decreasing luminosity towards larger linear sizes.

\subsubsection{The relationship between radio luminosity and jet power}

\begin{figure}
    \centering
    \includegraphics[width=1\linewidth]{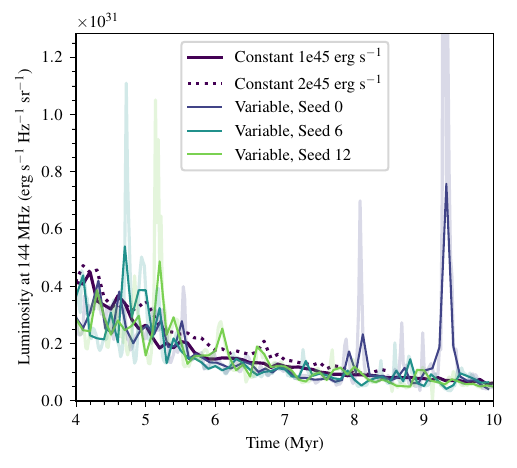}
    \caption{Luminosities at 144 MHz for variable and steady jets, with variable jets resampled to match frequency of steady jet data saves. The $2\times 10^{45}$ erg s$^{-1}$ constant power jet outgrows the simulation domain at $t\approx8.5$ Myr and has a slightly higher luminosity than the $10^{45}$ erg s$^{-1}$ constant power jet. The variable power jets have a similar radiative efficiency to the constant jets at most times, with some short periods of much higher radiative efficiency.}
    \label{fig:resampled_luminosities}
\end{figure}

Motivated by our findings of large amounts of variability in hotspot luminosity for modest changes in jet power, we continue by considering jet power estimates. A key component of this estimate is the radiative efficiency of the system, defined as the ratio of radio luminosity ($L_{\rm radio}$) to jet power~\citep[e.g.][]{Godfrey2013AgnLobes}. To investigate the differences in radiative efficiency over time, we show the luminosities of the two steady and three variable jets across their lifetimes in Figure~\ref{fig:resampled_luminosities}. Here we resample the data for the variable jets, showing only every fifth measurement to match the frequency with which we save the steady power jet luminosities. We include the luminosities without resampling for reference. For the majority of the variable jets' lifetimes their luminosities are comparable to that of the constant $10^{45}$ erg s$^{-1}$ steady jet, implying that their radiative efficiency is in general very similar to that of a steady jet. However, superimposed on this behaviour are the large spikes in luminosity previously discussed, together with smaller dips in luminosity. During the spikes in luminosity, the instantaneous radiative efficiency of the $\sigma=0.33$ variable jets is much higher. Methods for estimating instantaneous jet power include calculations based on the luminosity and size of the hotspot~\citep[e.g.][]{Godfrey2013AgnLobes}, and the use of scaling relations with luminosity at a particular wavelength, for example~\citep{Willott1999TheSurvey}. Crucially, the effects of interacting shocks and/or variable jet powers are not considered by these models. We therefore highlight the importance of being able to understand whether the jets we observe have been in low or high power periods in their recent history, in order to provide vital context for any energy estimates.

Given the above, signatures of variability can provide vital context for energy and power estimates. We have shown that a jump in jet power $\Delta Q_j\lesssim \langle Q_j\rangle$ can lead to two kinds of features in our computed emissivity maps. Firstly we have shown that the resulting travelling shock can cause a structure in the emissivity map which appears elongated in a direction perpendicular to the jet axis. Given an estimate of the inclination angle, as discussed in Section~\ref{sec:step_change}, the identification of this structure in a radio galaxy would give a lower limit on the time since the jump in power. On the one hand, travelling shock structures will in general be challenging features to pick out in most radio images; they can be relatively faint features, and are transient, so would only be expected to be present in a small fraction of sources. On the other, the travelling shock features we see also bear some resemblance -- in terms of physics and appearance -- to so-called double-double radio galaxies or restarting sources~\citep[e.g.][]{2000SchoenmakersRadioNuclei,2007BrocksoppThreeGalaxy,2019MahatmaLoTSSField}, and one can envisage a continuum of variability properties that produce a variety of travelling shock structures.

The increased hotspot luminosities that we see have lifetimes on the order of a few tenths of a Myr. If we were able to tell that a particularly bright hotspot were a consequence of flickering, then the observation of that bright hotspot would tell us that the increase in jet power must have happened within a jet travel time plus a few tenths of a Myr ago. The major challenge here is in discerning that a bright hotspot was indeed caused by flickering. Population-based approaches may help to mitigate these challenges and these are discussed in the following section. Furthermore, the level of brightening expected due to the interacting shocks mechanism can be very high, with increases of $\approx10\times$ in some cases (see Figure~\ref{fig:luminosity_vs_time}). Further work should compare these expectations to those from interactions of a jet with a sudden change in the ambient density field. Whilst we cannot hope to monitor a single source on these timescales, further constraints on how bright we can expect a hotspot to get in various cases in comparison to other parts of a radio galaxy or in comparison to the hotspots of other radio galaxies may provide a path towards identifying whether the underlying cause is likely to be a recent change in jet power.

\subsubsection{Prospects for identifying flickering in radio surveys}
Probing variability using radio surveys constitutes an aspirational goal, which would require detailed methodologies outside of the scope of this paper. In this section, we aim to provide some preliminary ideas for how this goal might be achieved.

The morphological features we identify depend on the variability of the simulated source over the last fraction of a Myr. This timescale implies that population studies would be a requirement for improving understanding of variability across AGN jet lifetimes, as we could not expect to sample the full range of behaviours found in the simulations in a single observed source. Recent radio surveys~\citep[e.g.][]{Jarvis2016TheSurvey,McConnell2020TheResults,Shimwell2022TheRelease} have led to large increases in the numbers of radio galaxies observed. In particular, the LOFAR Two-metre Sky Survey includes images of deep fields at 0.3" resolution~\citep[e.g. ELAIS-N1;][]{DeJong2024IntoImages,Shimwell2025TheField}. Within these deep fields, many sources are spatially resolved enough to pick out hotspots and some bright emission along the jet lines. However, there is a much small number of observed sources with the required level of spatial resolution to allow detailed comparison with the finer detail morphological features we identify, such as the emission from particles accelerated in travelling shocks. Hercules A and Cygnus A~\citep[e.g.][]{Perley1984TheA} are two examples of radio galaxies observed at spatial resolutions that could allow this comparison. Hercules A features ring-like structures that are thought to be the result of multiple AGN outbursts~\citep[e.g.][]{Timmerman2022OriginObservations}. Our simulations show that these brighter regions persist only for a fraction of the source lifetime and therefore at a given time would only be present in a fraction of sources. Furthermore they are difficult to differentiate from other emission in a single snapshot, which is all we have access to for any given source. The wider availability of hotspot data motivates more detailed characterisation of timescales over which the hotspots brighten and fade in variable jets (from data such as those we present here) and further understanding of the exact characteristics of the variability that affect the luminosities reached in these simulations (analytic models, see Section~\ref{sec:step_change}). For example, one methodology would be to simulate further seeds, create a histogram of hotspot luminosities for jets within a given length range (e.g. $40-50$ kpc) and then compare this to a histogram of hotspot luminosities from real sources to see if the distributions are significantly different, or more similar than a distribution of luminosities given a constant power jet simulation. Comparison of such observed distributions with predictions from simulations may then provide a useful way to understand whether the variability predicted by fuelling mechanisms such as CCA is compatible with radio survey data.

Further insight into the connection between jet activity and AGN fuelling may come from combining radio data with information about the AGN host galaxy and environment. \cite{Gaspari2013ChaoticHoles} suggest that CCA may be more common in environments such as hot galactic halos, groups or clusters of galaxies. \cite{LylaJung2025OnJets} find that galaxies close to cosmic filaments tend to have their major axis aligned with the closest filament, whilst radio jets in these environments tend to be more randomly oriented with respect to the host galaxy major axis compared to those further from cosmic filaments. These authors suggest that the random alignment of the jets is likely to be a consequence of chaotic accretion due to frequent mergers along the cosmic filaments. Analogously, we suggest that the distributions of hotspot luminosities could be studied in galaxies found in different environments (e.g. galaxies that are within or outside of clusters) in order to ascertain whether galaxies in, for example, clusters have a distribution of luminosities which would fit that expected from the evolution of a variable jet. Future facilities such as the SKA are likely to increase sample sizes both of hotspot luminosities and sources with spatially resolved backflows and lobes. With a larger sample of spatially resolved sources, particularly with improvements to sensitivity, it may become more feasible to search for the signatures of variability such as travelling shocks. 

\subsection{Limitations and areas for future work}
\label{sec:limitations}

There are a number of potential areas for future investigation, which include variability on shorter timescales, explicit modelling of magnetic fields using RMHD and varying the properties of the ambient medium into which the jet propagates. Here we discuss expectations for how each of these could affect our results, together with some suggestions for future work including these effects.

As mentioned in section~\ref{sec:setup}, the shortest timescale of variability that we are able to include in these simulations is limited by numerical issues. These issues stem primarily from the increased variance in the Lorentz factor that accompanies the shorter timescales. Variability on shorter timescales would lead to a shorter initial separation between sections of the jet travelling at different speeds. The hierarchical merging of shocks as a dissipation mechanism has been suggested in the context of x-ray binary hard state jets~\citep{Malzac2014TheShocks}. For our shortest timescale of variability (100 kyr) the distance over which two neighbouring shells of material merge depends heavily on their individual Lorentz factors. Taking an illustrative example of two shells with Lorentz factors $\Gamma_1=2.0$ and $\Gamma_2=2.5$, we get an initial separation of $\Delta d = 26$ kpc between the centres of the two regions in the observer's frame. The second shell has a velocity of $v_2'=0.245c$ in the frame of the first shell. The centres of the shells meet after 88 kyr in the observer's frame, during which time the centre of the first shell moves $d=23.0$ kpc in the observer's frame. These shells would therefore meet as the first shell reached the end of the jet in our simulations. Shorter timescales of variability would result in shocks merging at smaller distances from the jet base. A similar effect would be achieved by a larger $\Delta \Gamma = \Gamma_2-\Gamma_1$. At some lower limit of variability, almost all shells would merge prior to where our simulations begin, such that any variability below that timescale would be smoothed out. For example, taking the two shells with Lorentz factors $\Gamma_1=2.0$ and $\Gamma_2=2.5$ from the above discussion and shrinking the time between the centres of the two shells entering the simulation to 10 kyr implies they fully merge when the first shell has travelled a few kpc. 

In this work we chose not to explicitly model the magnetic fields for a variety of reasons. It is difficult to get observational constraints on the geometry of the magnetic field, which motivates an approach investigating various geometries. In addition, by initially separating investigation of these two effects, we aim to understand the role of each and the effects of interplay between them more comprehensively. We expect this interplay to have interesting and important effects. As discussed in section~\ref{sec:step_change}, there is a complex relationship between the magnitude of the pressure jumps and the recent history of the jet power. The addition of explicitly modelled magnetic fields is likely to further increase the complexity of this connection. Furthermore, magnetic fields may offer some shielding from the onset of instabilities via magnetic tension or, alternatively, lead to further instabilities via $m=1$ kink modes in regions with strong toroidal magnetic fields~\citep[e.g.][]{Upreti2024BridgingPolarisation}.

It is worth considering other potential mechanisms which could lead to similar brightened hotspot signatures. In particular we might expect intermittent hotspot brightening in the case of a clumpy ambient medium, for example in the case of a galaxy cluster. At smaller scales (within a few kpc from the base of the jet), the interaction of the jet beam with a large ($\gtrsim50$ pc)  cloud has been shown to decelerate the jet~\citep[e.g.][]{Mukherjee2025Jet-FeedbackReview}. Radio observations of the radio galaxy 3C40B in the cluster A194 show interactions between a radio jet and an intracluster magnetic filament~\citep{Rudnick2022IntraclusterJet}. Here the jet appears to change direction at the point at which it interacts with the filament, with the filament appearing to bend around this change in jet direction. There is some evidence for particle acceleration, potentially including acceleration up to X-ray energies at the meeting point of these structures, although random coincidence with a background X-ray source is not ruled out. Future work could investigate to what extent enhanced ram pressure can drive brightened hotspots in clumpy environments on larger scales, together with quantifying the amount of brightening expected in these situations. Disentangling the various potential causes of hotspot brightening in a source is likely to prove challenging, but provides strong motivation for the semi-analytical model of the amount of brightening expected in various situations (interacting shocks as compared to interactions with ambient medium), as mentioned in Section~\ref{sec:step_change}.

We simulated a radio galaxy at $z=0$, where inverse Compton scattering is a subdominant effect (equivalent to a field of $B_{\rm eq}\approx 3\,\rm{\mu G}$ at $z\approx0$); however, it can start to become a significant source of cooling relatively quickly with increasing redshift, $z$, with the energy density of the CMB, $u_{\rm CMB}=\frac{4.0\sigma_{B}}{c}T^4_{0}(1+z)^4$~\citep{Vaidya2018AFlows}, where $T_0= 2.728$ K and $\sigma_{B}$ is the Stefan-Boltzmann constant. The energy loss rate, $\dot{E}$ is proportional to $u_B+u_{\rm CMB}$. Taking a characteristic magnetic field strength $B=20\rm{\,\mu G}$ from a late time in one of our simulations (see Figure~\ref{fig:estimatedBfield}), the CMB provides an equivalent magnetic field, $B_{\rm eq} = 20\rm{\,\mu G}$ at a redshift of $z\approx2.5$. For observations at a single frequency the first order effect of a higher redshift will therefore be on the normalisation of the emissivity at each point in the source. However, some interplay with synchrotron cooling rates could also lead to changes in relative emissivities between regions in a source.

\section{Conclusions}
\label{sec:conclusions}

We have presented high resolution simulated radio emission maps for steady and flickering AGN jets on scales up to $60$ kpc. These simulations model the jet propagation using high resolution RHD, evolving energy spectra for populations of electrons as they move around the simulation domain. This method accounts for particle acceleration and cooling dependent on the conditions local to each population of electrons. Moreover, we have introduced the use of a fluid tracer to improve the modelling of adiabatic cooling in the case of a high density contrast between the injected jet and the ambient medium, removing spurious adiabatic heating in the case of mixing with unshocked material. Our methods allow improved modelling of emission along the jet line and at the hotspot, but underestimates diffuse emission in the lobes.

Our main conclusions are as follows:
\begin{enumerate}
    \item Our simulations show that a rapid increase in jet power leads to a dramatic increase in hotspot luminosity. This increase is intrinsically connected to the variability -- the same increased hotspot luminosity is not seen in a steady jet with higher power.
    \item As a consequence, in a variable jet, a particularly bright hotspot compared to other regions along the jet may suggest a recent period of high jet power and by extension a previous episode of high accretion rate. This suggests that spatially resolved observations of hotspots could be an effective probe of accretion history over the previous $\approx0.2$ Myr if further work can separate the effects from those of interactions with the surroundings, for example through the expected magnitude of brightening in each case.
    \item In our simulations, the sharp increase in hotspot luminosity is caused by the merging of a shock that travels along the jet with the pre-existing forward and reverse shock structure at the jet head. Whilst the time-averaged radiative efficiencies of the simulated constant power and variable jets are similar, the sharp increases in hotspot luminosity in the variable jets lead to a dramatic but short-lived increase in the radiative efficiency of the jet. This implies that should we accept that jets may vary in this way, jet power estimates based on hotspot luminosities should account for the interacting shock mechanism.
    \item The travelling shock can also lead to a bright region which travels along the jet axis. If observations were spatially resolved along the full jet axis, this could potentially be used to estimate a maximum time since an increase in power occurred, allowing the recent accretion history to be probed in further detail.
\end{enumerate}

Our results have significant implications for interpreting observations of radio galaxies. We have shown that variable jets produce a wide range of morphologies and that a single variable jet can move between these states multiple times in just a short fraction of its overall lifetime. Furthermore, we establish that the morphology of a source can have a strong dependence on the underlying fuelling mechanism through the effects of variability in jet power. Indeed, a variable jet may well show a very bright hotspot that is a result of a high power period which is no longer observable in the core region due to the jet travel time. Variable jets may also produce relatively bright patches of emission from shocks which travel along the jet immediately following increases in jet power, however these are likely to be very challenging to observe, particularly with current and near future radio telescopes. Intermittent hotspot activity in jets with variable powers challenges the picture of a smooth evolution of luminosity with age, which is often used in the interpretation of population studies. In the future, a combination of statistical constraints from surveys of radio galaxies and statistical predictions from simulations such as these could be used to test the predictions of fuelling theories such as CCA.

\section*{Acknowledgements}
JHM and ELE acknowledge funding from a Royal Society University Research Fellowship (URF\textbackslash R1\textbackslash221062). The authors acknowledge the use of resources provided by the Isambard 3 Tier-2 HPC Facility. Isambard 3 is hosted by the University of Bristol and operated by the GW4 Alliance (https://gw4.ac.uk) and is funded by UK Research and Innovation; and the Engineering and Physical Sciences Research Council [EP/X039137/1]. We are grateful for the use of the following software packages: PLUTO~\citep{Mignone2007PLUTO:Astrophysics}, matplotlib~\citep{Hunter2007Matplotlib:Environment}, pandas~\citep{McKinney2010DataPython,Thepandasdevelopmentteam2024Pandas-dev/pandas:Pandas}, scipy~\citep{Virtanen2020SciPyPython}. The authors wish to thank the anonymous referee for their helpful and constructive comments which have improved the quality of this manuscript.

\section*{Data Availability}
 
The data underlying this article are available from the authors on reasonable request.



\bibliographystyle{mnras}
\bibliography{references1} 




\appendix

\section{Estimated Magnetic Field}
\label{sec:app_estimated_magnetic_field}
\begin{figure}
    \centering
    \includegraphics[width=\linewidth]{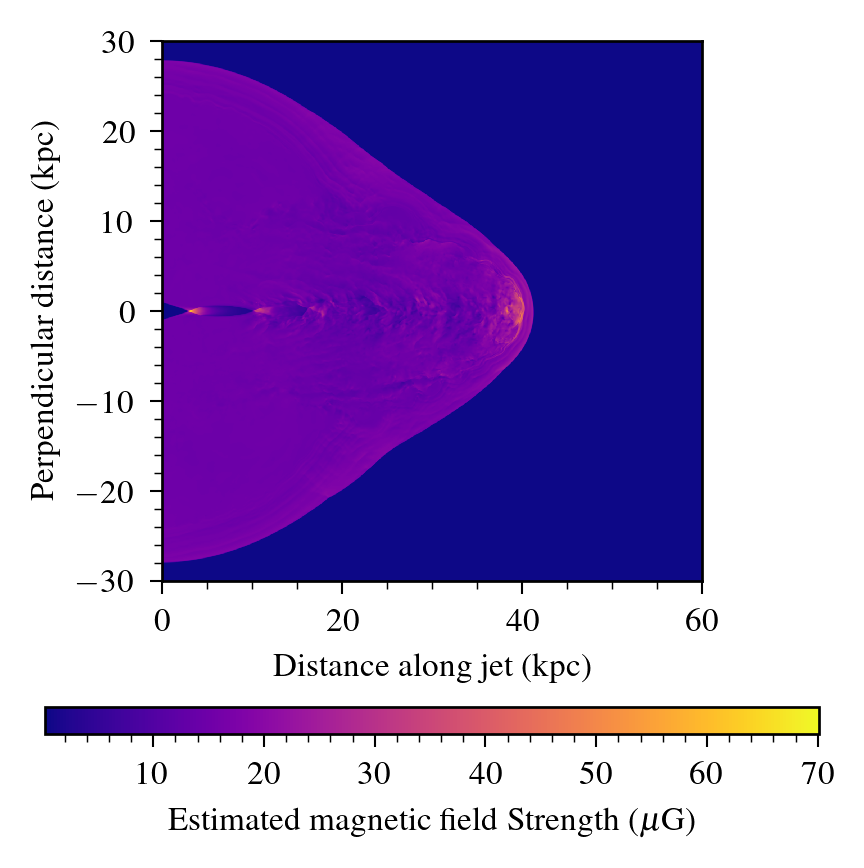}
    \caption{Estimated magnetic field for a slice through the variable jet with Seed 12 at 9 Myr. The estimated magnetic field strengths in the hotspot and cocoon areas are approximately $50 \rm{\,\mu G}$ and $10-20 \rm{\,\mu G}$ respectively.}
    \label{fig:estimatedBfield}
\end{figure}

We present an example of the estimated magnetic field strength used in our simulations, as calculated based on the internal energy density of the fluid (see Section~\ref{sec:shock_acceleration}). Figure~\ref{fig:estimatedBfield} shows the estimated magnetic field strength in a snapshot of the variable power jet simulation with seed 12 at 9 Myr. The hotspot has an estimated magnetic field strength of around $50 \rm{\mu G}$ at this time, whereas that of the cocoon is approximately $20 \rm{\mu G}$. At earlier times in the simulation the strength of the magnetic field in the lobes is higher. Because the pressure at the hotspot in our simulations is greatly variable, the estimated strength of the magnetic field at the hotspot will also vary greatly throughout our simulations.

For illustrative purposes we compare the lobe magnetic field strengths in this snapshot with observational estimates of a source with similar size. 3C 438 has an estimated size of 98.3 kpc~\citep{Laing1983BrightGalaxies}, which is reasonably close to the size of our source accounting for both jets and assuming near symmetry. Field strengths in the range of $36.5\rm{\,\mu G}$ to $40.2\rm{\,\mu G}$ were found for this radio galaxy~\citep{Harwood2015SpectralGalaxies} through fitting of the Kardashev~\citep{Kardashev1962NonstationarinessEmission} and Pacholczyk~\citep{Pacholczyk1971RadioSources} (KP) and Jaffe-Perola~\citep{Jaffe1973DynamicalGalaxies} (JP) spectral ageing models respectively. Slightly lower magnetic field strengths were found when the jet region was excluded from the fitting. These models assume a fixed magnetic field across the source, which is not a realistic assumption around the hotspot of the source, but give us some idea of the strengths we might expect in the lobes. These strengths are reasonably close to those obtained in our simulations, suggesting that our methods can reproduce reasonable estimates for magnetic field strengths in the lobes of radio galaxies.
\section{Particle injection}
\label{sec:app_particle_injection}

Here we describe our approach to injecting Lagrangian particles into the simulation. Lagrangian particles are initialised at the base of the jet at each time step in order to sample the jet material. The number of particles injected and the frequency with which they are injected is a decision that calls for a balance between reaching an adequate sampling of the fluid and the amount of data created each time the particles files are written. The energy injection algorithm is such that if two particles are in the same computational cell at a shock, the energy assigned to the non-thermal electrons represented to each Lagrangian particle (see Section~\ref{sec:shock_acceleration}) is half that which would be given to a single Lagrangian particle if it were the only one present. Ideally, therefore, we would aim for at least one particle per cell in the simulation, where above this, the emissivity predicted will not increase, but any fewer particles results in a reduced emissivity prediction. In reality, injecting one particle per cell does not lead to having exactly one particle per cell throughout the simulation, as particles collect in regions with a high density of jet material.

The data created by simulations with one particle per cell would prevent us sampling the jets at the time frequency we use in this study. Therefore, to reduce the computational cost and data storage requirements of our simulations, we inject less than one particle per cell and correct for the undersampling. We inject particles uniformly across the jet base and stagger their precise locations across timesteps. We inject 156 particles every 3 timesteps. The jet has an area of $25^2\pi \approx 1963$ cells in our simulations. With a maximum Courant number of 0.3 we would need to inject 589 particles per timestep to have one Lagrangian particle per cell on average. We therefore calculate a correction factor of approximately 11.5 for the emissivity. To test that this is reasonable, we ran a short section of the $10^{45}$ erg s$^{-1}$ constant power simulation with the increased number of particles and compared the emissivities from the same point in the subsampled simulation. We also tested that the morphology predictions did not change significantly with this level of undersampling by creating subsampled maps from different random cuts of the particles in the test simulation.

\section{Adiabatic cooling with mixing test cases}
\label{sec:app_adiabatic_cooling}

\begin{figure*}
    \centering
    \includegraphics[width=\linewidth]{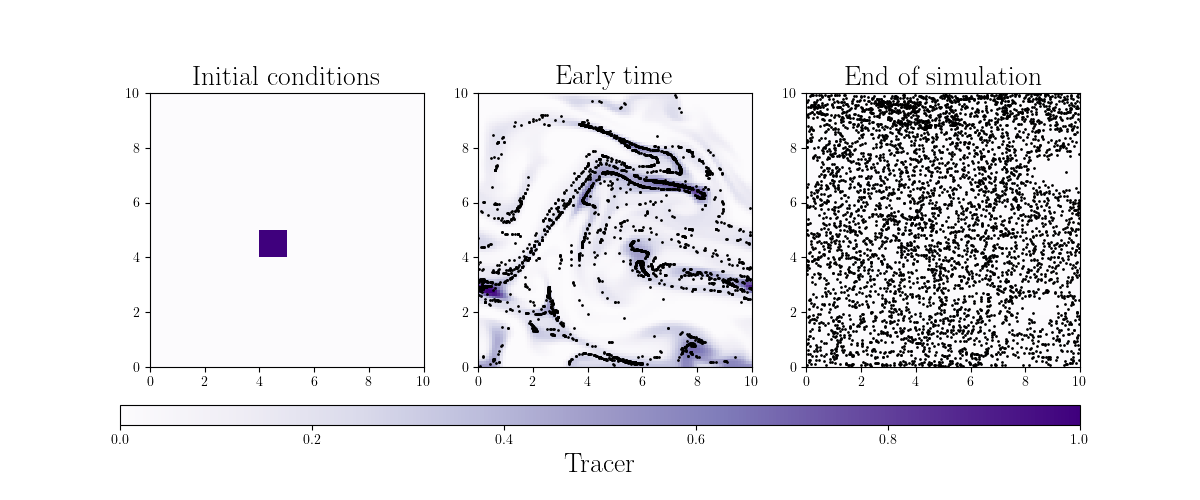}
    \caption{The shocked population mixes with the surrounding material (here of equal density), resulting in a near uniform distribution of fluid tracer and Lagrangian particles.}
    \label{fig:equal_density_tracer}
\end{figure*}

We present two tests run during the development of our method to include a factor of tracer in the adiabatic cooling term as discussed in Section~\ref{sec:tracer_method_derivation}. These tests highlight the need for an improved treatment of mixing between shocked and unshocked material. In the first, hereafter test A, a portion of shocked material is mixed into a much larger region of unshocked material with the same density. In the second, hereafter test B, we again mix a portion of shocked material into a much larger region of unshocked material, but here the shocked material has a density which is $1/10$ of that of the surrounding material. This case reflects the physical situation we model in our simulations of AGN jets, as we see mixing of low density jet material with comparatively much higher density ambient medium material. In both cases we initialise a population of Lagrangian particles in the region of shocked material, giving each Lagrangian particle a power law electron energy spectrum. We initialise the fluid velocity with alternating positive and negative values along one axis in order to cause mixing via Kelvin Helmholtz instabilities. Figure~\ref{fig:equal_density_tracer} illustrates this setup for the equal density case, showing the initial location of the tracer material (1/100 of the simulation volume) and the subsequent mixing with the other material in the simulation. The Lagrangian particles are not shown in the first panel as they would obscure the location of the fluid tracer; they begin uniformly spaced within this region. Both the fluid tracer and the Lagrangian particles are spread throughout the volume by the end of the test simulation, showing that mixing has occurred between the shocked and unshocked material. For the purposes of these tests, we turn off synchrotron and inverse Compton cooling, and allow only adiabatic cooling and heating.

\begin{figure}
    \centering
    \includegraphics[width=\linewidth]{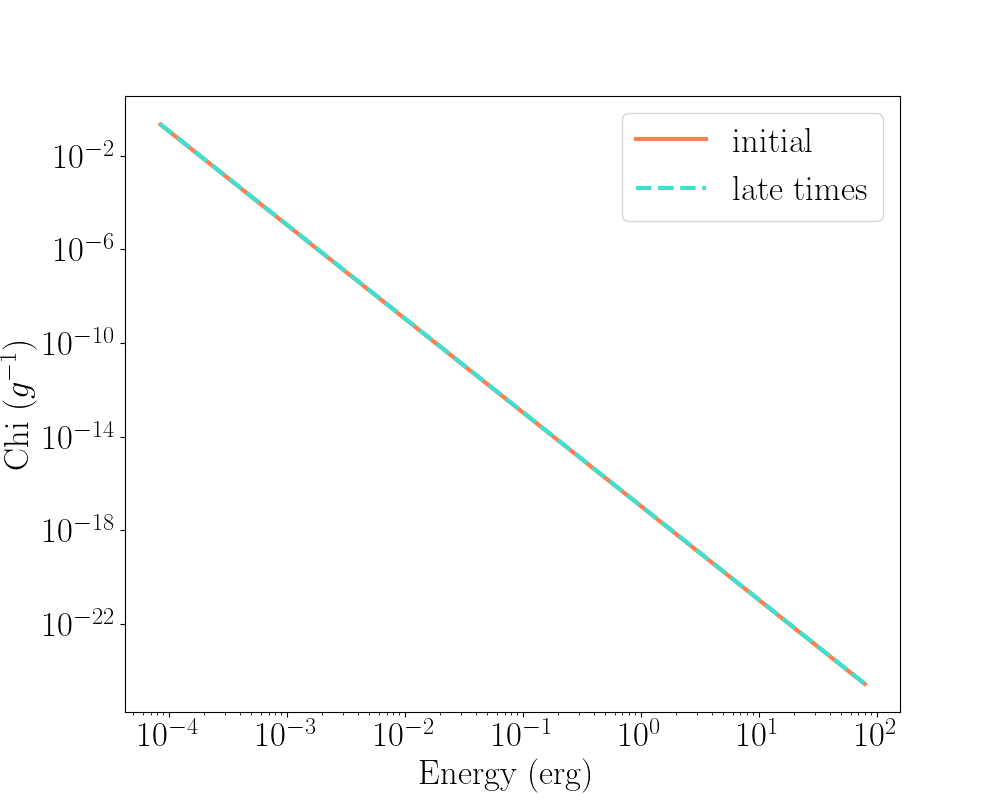}
    \caption{Test A, original scheme. Distribution of $\chi(E) = \mathcal{N}(E)/n$ in the case where a shocked population mixes with an equal density ambient medium following the original scheme. $\chi$ is shown for a randomly picked Lagrangian particle at the start and end of the test simulation.}
    \label{fig:equal_density_chis}
\end{figure}

\begin{figure}
    \centering
    \includegraphics[width=\linewidth]{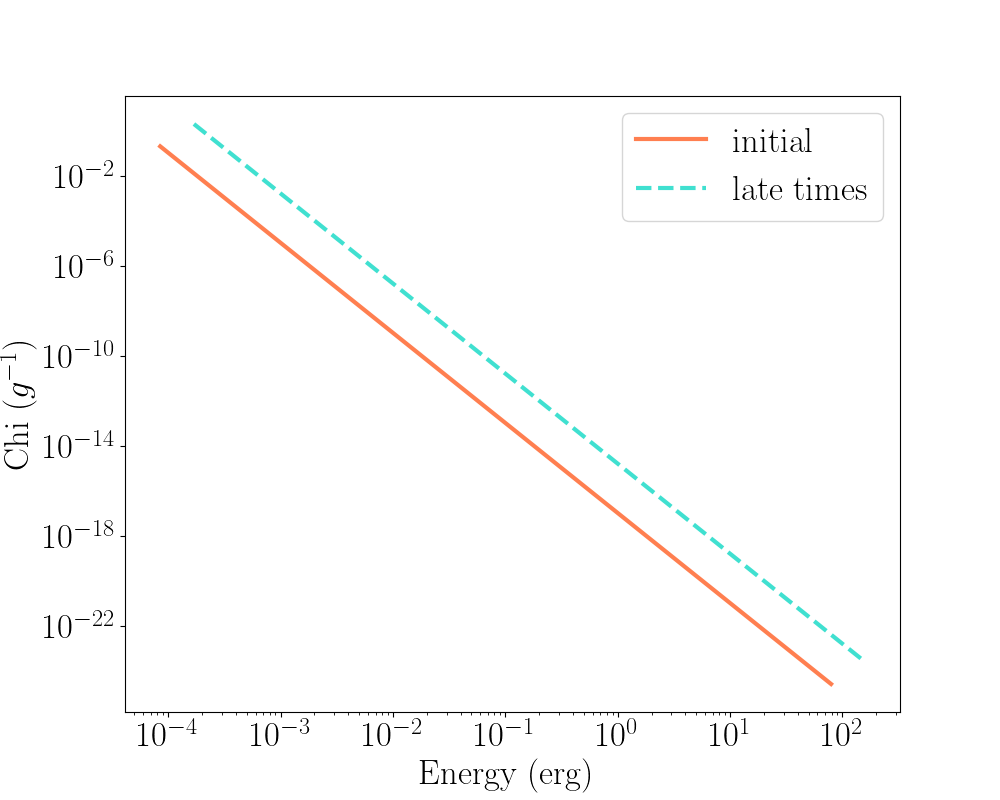}
    \caption{Test B, original scheme. Distribution of $\chi(E) = \mathcal{N(E)}/n$ with energy in the case where a shocked population mixes with a denser ambient medium following the original scheme. $\chi$ is shown for a randomly picked Lagrangian particle at the start and end of the test simulation.}
    \label{fig:underdense_tracer_chis}
\end{figure}

We now present the results of test A when run using the original Lagrangian particle module~\citep{Vaidya2018AFlows}. In the original Lagrangian particle module the quantity $\chi(E)$ is normalised to the number density of all fluids (shocked or unshocked) such that $\chi(E) = \mathcal{N}(E)/n$. This normalisation is used in Figures~\ref{fig:equal_density_chis} and~\ref{fig:underdense_tracer_chis} which are discussed below. Figure~\ref{fig:equal_density_chis} shows the distribution of electron energies at the start and end of test A (mixing into an equal density medium). As expected based on the original scheme, the value of $\chi$ remains constant at each electron energy. Over the time the test case is evolved for, the number density of electrons, $n$, stays constant throughout the simulation volume. Because $\chi$ and $n$ remain constant, the prediction for $\mathcal{N}$, the  number density of shocked electrons, also remains constant. However, as the shocked material now fills the full volume, rather than just the small dark area shown in the first panel of Figure~\ref{fig:equal_density_tracer}, physically (neglecting for now any work done by the shocked fluid on surrounding fluid) we expect $\mathcal{N}$ to instead now have $1/100$ of its original value. In the case of mixing, some improvement to the scheme is therefore required to avoid overestimating the number density (and therefore the emissivity) of shocked material.

The motivation for an improved scheme is strengthened in the case of mixing of shocked material into denser surroundings. We therefore now present the results of test B as run with the original module in Figure~\ref{fig:underdense_tracer_chis}. Across the time the test covers $\chi(E) = \mathcal{N}(E)/n$ increases for each electron energy and the energy bins also move to higher energies. The value of $n$ as sampled by the Lagrangian particles is almost $10\times$ higher at the end of the test, compared to at the beginning. This means that $\mathcal{N}(E)$, the prediction of the number density of shocked electrons, has increased in this case, which is the opposite of what we expect to happen in the physical case.

\begin{figure}
    \centering
    \includegraphics[width=\linewidth]{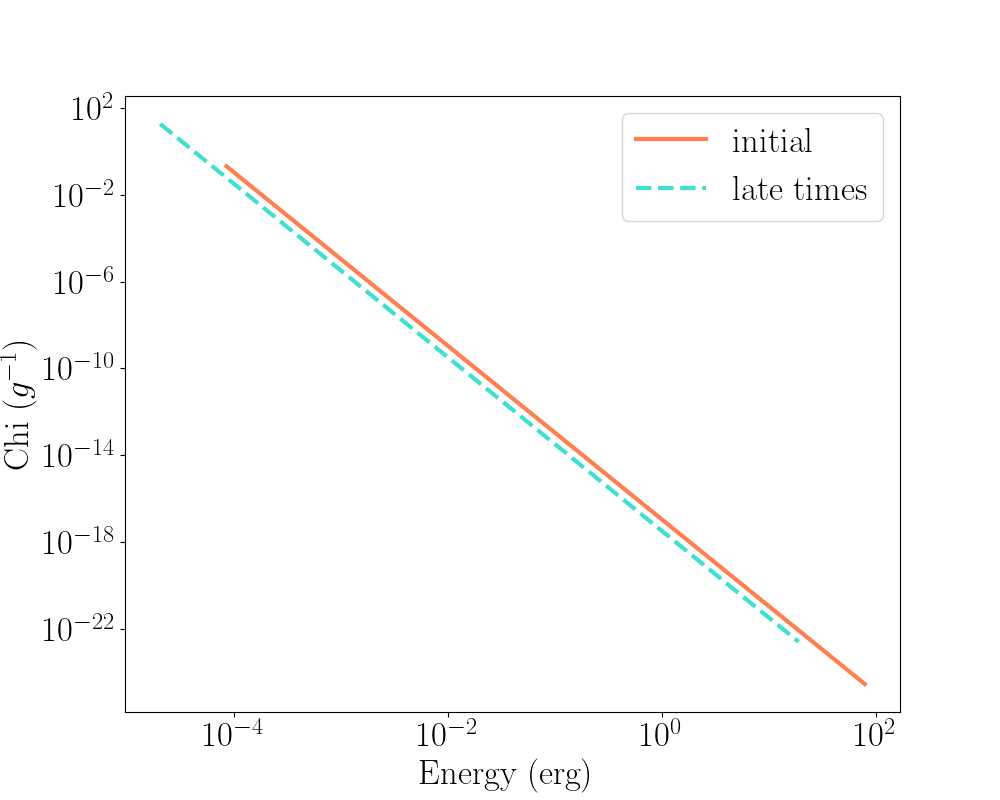}
    \caption{Test A, updated scheme. Distribution of $\chi(E) = \mathcal{N(E)}/(\mathcal{C}n)$ with energy in the case where a shocked population mixes with an equal density ambient medium following the original scheme. $\chi$ is shown for a randomly picked Lagrangian particle at the start and end of the test simulation.}
    \label{fig:updated_scheme_equal_density_chis}
\end{figure}

We now turn to the results of mixing tests for our updated scheme (as described in Section~\ref{sec:tracer_method_derivation}) for test A -- the case with equal densities for the initial shocked and unshocked fluids. We return to defining $\chi(E) = \mathcal{N}(E)/(\mathcal{C}n)$, where $\mathcal{C}$ is the value of the fluid tracer in the simulation. At the locations of the tracer particles, $\mathcal{C}$ should decrease to approximately 1/100 of its original value over the course of the test simulation. The number density of electrons $n$ should remain constant, and so we expect the number density of the shocked particles to reduce to 1/100 of its original value. We therefore expect the value of $\chi$ to remain approximately constant, but for the distribution to move to slightly lower energies due to work done by the shocked electron populations to expand. As the distribution moves to lower energies, the number of particles per unit energy must increase to preserve the overall number of particles. Figure~\ref{fig:updated_scheme_equal_density_chis} shows the results of this test. As expected, the distribution has moved to lower energies and higher number densities of particles per unit shocked fluid density. Taking the maximum energy bin as an example, following Equation~\ref{eqn:bin_edge_evolution}, we expect the energy to change to $E_{\rm max}(t)\approx E_{\rm max}(t_0)(\frac{1}{100})^\frac{1}{3}\approx0.2E_{\rm max}(t_0)$, which is approximately what is seen in Figure~\ref{fig:updated_scheme_equal_density_chis}.

In this discussion we have largely neglected discussion of any work done by the fluid in order to expand into its surroundings. This should only lead to a further decrease in the energies of the particles and so does not affect the above statements that we expect $\chi$ to decrease in each case. Our use of an adiabatic scheme, passive tracers and tracer particles, involves a number of intrinsic assumptions regarding the interaction between the nonthermal electrons and the thermal electrons. For example, we do not consider heat transfer between the two populations, modification of the plasma equation of state by the nonthermal population, or `microphysical' plasma interactions or instabilities on, e.g., the gyroradius scale of the particles. The assumption of no heat transfer will be less valid in the case of mixing populations, however the major effect in the case of expansion from the hotspot region of a jet is that of a reduction in concentration of the shocked material, such that overall this approach greatly improves the modelling of adiabatic expansion for the shocked fluid compared to not including the tracer factor.


\section{Particles vs fluid tracer}
\label{sec:particles_fluid_tracer}
\begin{figure}
    \centering
    \includegraphics[width=\linewidth]{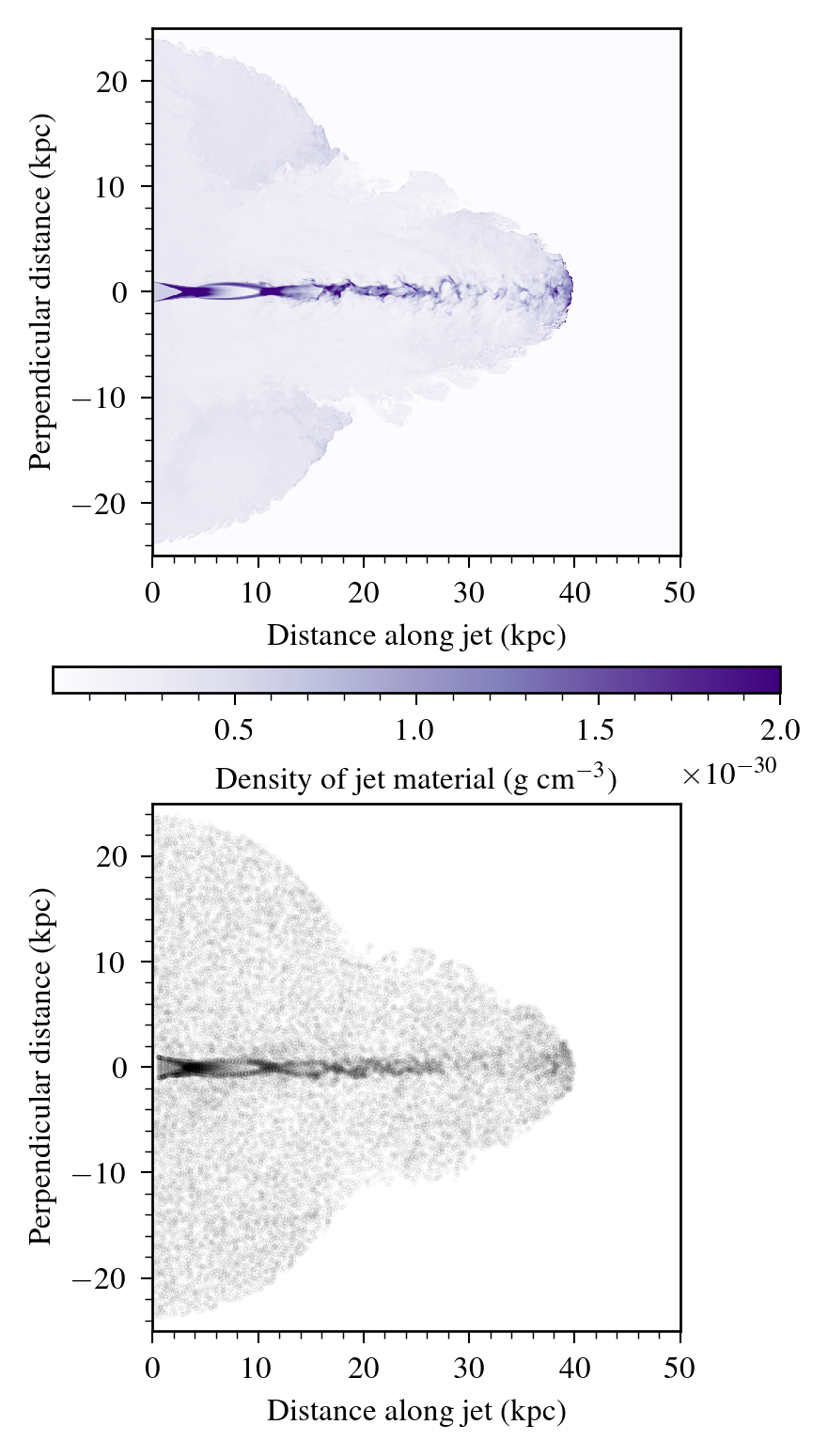}
    \caption{Seed 12, 9 Myr snapshot of fluid tracer multiplied by density and location of Lagrangian particles in slice through centre of simulation domain. The distributions of Lagrangian particles show good agreement on medium and large scales.}
    \label{fig:tracer_comparison}
\end{figure}

\begin{figure}
    \centering
    \includegraphics[width=\linewidth]{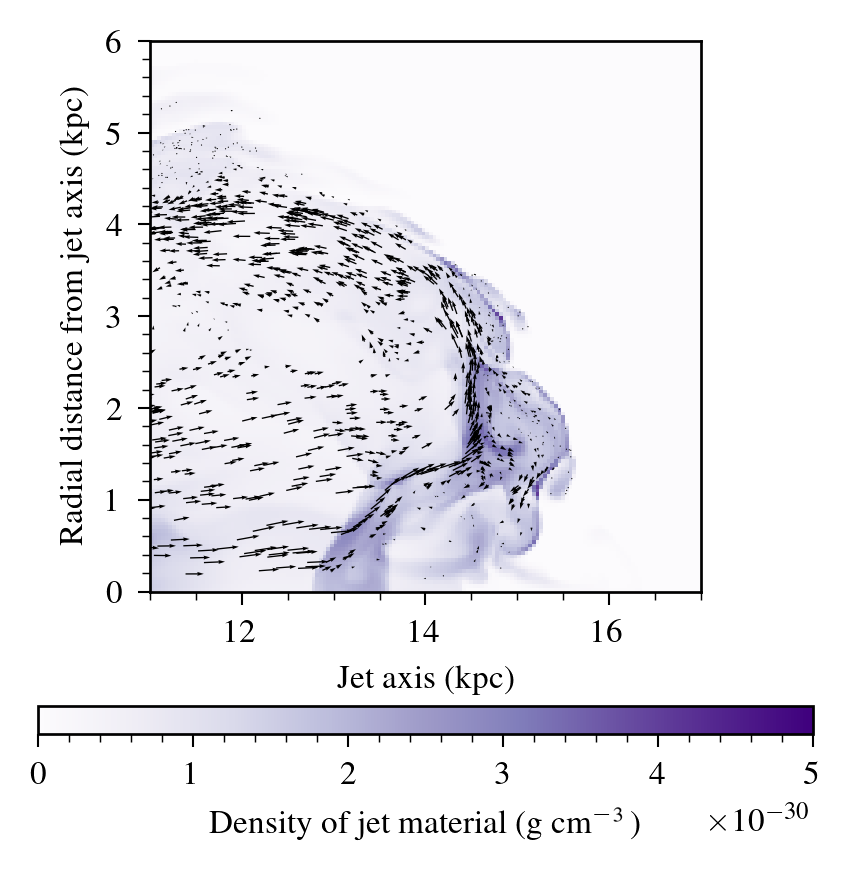}
    \caption{The velocity and distribution of Lagrangian particles overlaid on the fluid tracer multiplied by density in a 2D axisymmetric simulation used for testing. We show the jet head in this image with the jet moving from left to right. As this simulation is axisymmetric, we show only one half of the jet head here. This simulation has the same resolution of 0.04 kpc (25 cells per jet radius) as our 3D simulations and illustrates that in the case of strong backflows, this resolution is high enough to ensure that the Lagrangian particles effectively follow the backflows and enter the cocoon region.}
    \label{fig:tracer_particles}
\end{figure}

Figure~\ref{fig:tracer_comparison} shows the distribution of the scalar tracer multiplied by the fluid density and the distribution of Lagrangian particles for a slice through the seed 12 simulation at 9 Myr. This illustrates the agreement of the two tracer approaches in tracking the density of jet material. We found that the agreement of these two approaches was dependent on the resolution used for the simulations and that our resolution of 0.04 kpc (25 cells per jet radius) was sufficient to ensure this agreement. We found that a high resolution was particularly important near the jet head. We show the jet head region for a 2D test simulation at the same resolution in Figure~\ref{fig:tracer_particles}.
The locations of the Lagrangian particles in this simulation agree well with the distribution of the jet material. In particular, this test shows that when strong backflows form, the Lagrangian particles enter and move through these backflows effectively at this resolution. At lower resolutions we found that the Lagrangian particles often became trapped in the region between the forward and reverse shocks at the jet head.


\bsp	
\label{lastpage}
\end{document}